\documentclass[preprint, 12pt, nonatbib]{elsarticle}
\journal{journal}
\usepackage{xr}

\usepackage{array,bm,amsmath,amssymb,mathrsfs,epsfig,epstopdf}

\usepackage{setspace}
\usepackage{CJK}
\usepackage{color}
\usepackage{lscape}
\usepackage{booktabs}
\usepackage{verbatim}
\usepackage{enumerate}
\usepackage{threeparttable}
\usepackage{multicol}
\usepackage{longtable}
\usepackage{multirow}
\usepackage{bm}  
\usepackage{bbm}
\usepackage{caption}
\usepackage{algorithm}
\usepackage{algorithmic}
\usepackage{geometry}
\usepackage{framed}
\usepackage[pdftex,bookmarksnumbered,bookmarksopen,colorlinks, citecolor=blue,linkcolor=blue,urlcolor=blue]{hyperref}
\usepackage{graphicx}
\usepackage{amsmath}
\usepackage{amsfonts}
\usepackage{float}
\usepackage{ragged2e}
\usepackage[figuresright]{rotating}

\newcommand\Vector[1]{\bm{#1}}

\newcommand\vw{{\Vector{w}}}
\newcommand\vx{{\Vector{x}}}
\newcommand\vy{{\Vector{y}}}

\newcommand\MATRIX[1]{\bm{#1}}

\newcommand\mB{{\MATRIX{B}}}

\newcommand{\tabincell}[2]{\begin{tabular}{@{}#1@{}}#2\end{tabular}}

\makeatletter
\let\c@author\relax
\makeatother

\usepackage[backend=biber, minbibnames=1, maxnames=2, maxbibnames=2, sorting=none, citestyle=numeric-comp]{biblatex}
\begin{document}

\title{\bf A model-free feature extraction procedure for interval-valued time series prediction}

\author[1]{Wan Tian}
\ead{wantian61@foxmail.com}

\author[1,2]{Zhongfeng Qin\corref{cor1}}
\ead{qin@buaa.edu.cn}

\author[3]{Tao Hu}
\ead{hutaomath@foxmail.com}

\cortext[cor1]{Corresponding author}
\address[1]{School of Economics and Management, Beihang University, Beijing 100191, China}
\address[2]{Key Laboratory of Complex System Analysis, Management and Decision (Beihang University), Ministry of Education, Beijing 100191, China}
\address[3]{School of Mathematical Sciences, Capital Normal University, Beijing 100048, China}

\begin{abstract}
In this paper, we present a novel feature extraction procedure to predict interval-valued time series by combing transfer learning and imaging approaches. Initially, we represent interval-valued time series using a bivariate point-valued time series, which serves as a representative form. We first transform each time series into images by employing various imaging approaches such as recurrence plot, gramian angular summation/difference field, and Markov transition field, and construct an image dataset by treating each imaging method's output as a separate class. Based on this dataset, we train several candidates for a feature extraction network (FEN), specifically ResNet with varying layers. Then we choose the penultimate layer of the FEN to extract the most relevant features from the transformed images. We integrate the extracted features into conventional predictive models to formulate the corresponding prediction models. To formulate prediction, we integrate the extracted features into a regular prediction model. The proposed methods are evaluated based on the S\&P 500 index and three data-generating processes (DGPs), and the experimental results demonstrate a notable improvement in prediction performance compared to existing methods.

\end{abstract}

\maketitle	

\section{Introduction}

Modeling and forecasting of interval-valued time series has been gaining increasing attention in the fields of statistics and econometrics \cite{arroyo2006introducing, san2007imlp, maia2008forecasting, han2012autoregressive, gonzalez2013constrained, han2016vector}, as it plays a crucial role in capturing the uncertainty and variability of observed phenomena. In practical applications, interval-valued time series are quite common. For instance, in financial markets, we can use interval-valued time series to represent the range of highest and lowest prices of commodities within a trading period, providing a comprehensive view of price behaviors. Similarly, in meteorology, interval-valued time series are widely used to describe daily weather conditions, such as PM2.5 concentration and temperature. Moreover, in the medical field, interval-valued time series are frequently used to represent measurements like blood pressure. In general, modeling interval-valued time series has two main advantages over point-valued time series \cite{han2016vector}. Firstly, within the same time period, interval-valued time series contain more variation and level characteristics \cite{he2009impacts, gonzalez2013constrained, han2012autoregressive}, which means modeling based on interval-valued time series can lead to more efficient and robust prediction results. Secondly, specific disturbances that may act as noise in point-valued time series can be addressed by modeling interval-valued time series, thus mitigating the impact on inference.

The existing literature has extensively explored statistical and machine learning approaches for modeling and forecasting interval-valued time series \cite{arroyo2006introducing,san2007imlp, arroyo2007exponential,neto2008centre, arroyo2011different, han2012autoregressive, gonzalez2013constrained, han2016vector}. All these methods can be divided into two categories. The first category represents intervals using representative bivariate point values and processes them using point value-based time forecasting methods.
For instance, \textcite{san2007imlp} proposed the concept of an Interval Multilayer Perceptron (iMLP) to address interval-valued data. The iMLP is based on Interval Neural Networks (INN) and offers a comprehensive framework for modeling and analyzing interval-valued time series.   Another relevant study by \textcite{arroyo2007exponential} proposed an exponential smoothing method specifically designed for interval-valued time series. This method utilizes interval arithmetic \cite{moore1966interval} to capture and model the uncertainty inherent in the data. \textcite{gonzalez2013constrained} introduced a constrained regression model specifically designed for the two boundaries of the interval-valued time series . This model ensures that the natural order of the interval is preserved in all cases. The second category treats intervals as a whole and models them as a single entity. \textcite{han2012autoregressive} introduced the concept of extended random intervals and established the autoregressive conditional interval model for interval-valued time series, along with corresponding parameter estimation and statistical inference theory. Building upon the work of \textcite{han2012autoregressive}, \textcite{han2016vector} further developed the vector autoregressive moving average model for interval vector-valued time series. Additionally, \textcite{Sun2022} proposed a model averaging method for interval-valued time series, leveraging weight selection criteria.  The method involves combining multiple models to improve forecasting accuracy.

In recent years, the field of deep learning has witnessed significant advancements and has proven to be successful across various domains. The effectiveness of deep learning techniques is often attributed to their ability to learn meaningful representations directly from raw data \cite{Goodfellow-et-al-2016}. However, to the best of our knowledge, there is a limited number of studies that have explored the modeling and prediction of interval-valued data using modern deep learning techniques. The earlier references to interval neural networks (INN) primarily focus on single-layer perceptron models and do not fully leverage the representation power of deep learning models. Additionally, transfer learning has garnered significant attention in time series modeling. Transfer learning involves leveraging knowledge gained from training data in a related task to improve the performance of the target task \cite{ye2018novel, vercruyssen2017transfer, fawaz2018transfer}. However, these transfer learning studies have mainly focused on point-value time series modeling, and the source domains are not typically related to image-based data. As a nascent endeavor, our approach aims to model interval-valued time series data using deep learning techniques derived from computer vision \cite{he2016deep}, while also incorporating feature-based transfer learning. By exploring the potential of deep learning and transfer learning in the context of interval-valued time series, we aim to enhance the modeling and prediction capabilities and uncover new insights from the data.

In this paper, we propose a feature extraction procedure specifically designed for interval-valued time series. The procedure combines feature-based transfer learning and time series imaging, allowing for integration with regular prediction models for interval-valued time series forecasting. Our work builds upon the research conducted by \textcite{tian2021drought} in the context of point-valued drought time series, extending their approach to interval-valued time series. To characterize interval-valued time series, we employ central and range bivariate point-valued time series representation. These representations are divided into disjoint segments of a specific length, which are then transformed into images using the four imaging approaches described in \cite{eckmann1995recurrence, wang2015encoding}. By treating the images obtained from each imaging approach as distinct classes, we construct an image dataset comprising four classes through the aforementioned procedure. Subsequently, we train a selection of ResNet models \cite{he2016deep}, with varying layers, on the image dataset. These trained models, referred to as the feature extraction network (FEN), are then used to retrospectively extract features from interval-valued time series data. Furthermore, we conduct extensive experiments using both real-world and simulated data to evaluate the performance of the proposed feature extraction procedure. We compare the prediction performance of various regular prediction models before and after applying the feature extraction procedure.

The structure of this paper is as follows. In Section \ref{sec2}, we introduce the necessary symbols and formalize the framework for interval vector-valued time series prediction. Section \ref{sec3} presents the methodology proposed in this paper. It includes a detailed explanation of the construction of the FEN and the feature extraction procedure, which incorporates transfer learning techniques. The numerical results obtained from applying the proposed feature extraction procedure to real data and various DGPs are presented in Section \ref{sec4}. Finally, in Section \ref{sec5}, we conclude the paper by summarizing the key findings and contributions of our research.  We also discuss the potential implications and future directions for further exploration in the field of interval-valued time series prediction.

\section{Problem setup} \label{sec2}

Stochastic interval-valued time series is a sequence of interval-valued random variables indexed by time \(t\). The interval-valued time series \(\left\{y_t\right\}^T_{t=1}\) can be represented by lower and upper bounds, i.e., \(y_t = [y^l_t, y^u_t]\), or equivalently by using the center and range, i.e., \(y_t = (y^c_t, y^r_t)\), where \(y^c_t = \frac{1}{2}(y^l_t+ y^u_t), y^r_t = \frac{1}{2}(y^u_t-y^l_t)\). The restriction \(y^l_t \leq  y^u_t\) (or \(y^r_t \geq 0\)) ensures a natural order between the upper and lower bounds of the interval.

In this paper, we focus on modeling the dynamics of the process \(\left\{y_t\right\}^T_{t=1} = \left\{\left[y^l_{t}, y^u_{t}\right]\right\}^T_{t=1}\) within the framework of an information set that only includes the past history of the process, denoted as \(\vy_{t-1}  = \left(y_1,y_2,\cdots, y_{t-1}\right)\), without the inclusion of any exogenous variables. To remove the constraint of lower bound being smaller than the upper bound during the modeling process, we propose considering the joint dynamics of the central \(\left\{y^c_{t}\right\}^T_{t=1}\) and the range time series \(\left\{y^r_{t}\right\}^T_{t=1}\),
\begin{equation}\label{datapromodel}
y_t\overset{d}{=} \left[
\begin{array}{c}
y^c_{t}\\
y^r_{t}
\end{array}
\right] = \left[
\begin{array}{c}
M_l\left(\text{FEN}(\vy_{t-1}^c, \vy_{t-1}^r);\mB_l\right)\\
M_u\left(\text{FEN}(\vy_{t-1}^c, \vy_{t-1}^r);\mB_u\right)
\end{array}
\right] + \left[
\begin{array}{c}
\epsilon^c_{t}\\
\epsilon^r_{t}
\end{array}
\right],
\end{equation}
where \(\vy_{t-1}^c = \left(y^c_{t-1}, y^c_{t-2}, \cdots, y^c_1\right)\) and \(\vy_{t-1}^r = \left(y^r_{t-1}, y^r_{t-2}, \cdots, y^r_1\right)\) are the information sets corresponding to the centre and range of  timestamp \(t\), respectively, with \(\epsilon^c_{t}\) and \(\epsilon^r_{t}\) being the errors. By jointly modeling the center and range time series, we can recover the original interval-valued time series from the predicted values using simple arithmetic operations.

\begin{figure}[h]
\centering
\includegraphics[scale=0.36]{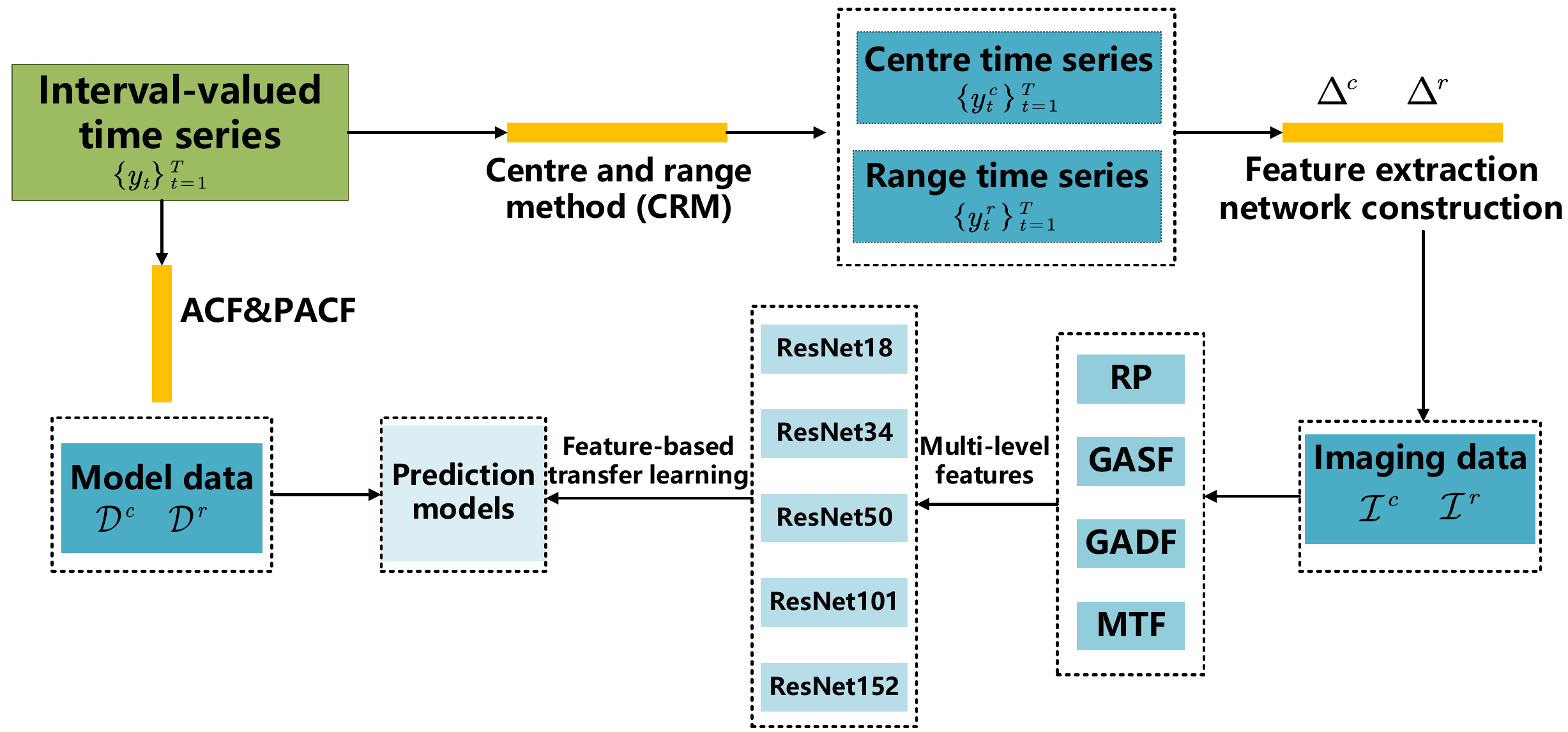}
\caption{\normalsize Technical roadmap for interval-valued time series prediction based on transfer learning and time series imaging.} \label{technicalroadmap}
\end{figure}

Figure \ref{technicalroadmap} shows the technical roadmap of this paper. In accordance with the model (\ref{datapromodel}), this section describes the construction of the Feature Extraction Network (FEN) and the feature extraction procedure based on the central time series \(\left\{y^c_{t}\right\}^T_{t=1}\) and the range time series \(\left\{y^r_{t}\right\}^T_{t=1}\). To determine the orders of the center time series \(\left\{y^c_{t}\right\}^T_{t=1}\) and the range time series \(\left\{y^r_{t}\right\}^T_{t=1}\), we can utilize the Autocorrelation Function (ACF) \cite{fan2008nonlinear} and Partial Autocorrelation Function (PACF) \cite{fan2008nonlinear}. By analyzing the ACF and PACF, we can determine the orders of \(a^c\) for the center time series and \(a^r\) for the range time series. Using the determined orders, we can construct the datasets \(\mathcal{D}^c\) and \(\mathcal{D}^r\) for interval-valued time series prediction. These datasets will be used to train the FEN and extract features from the interval-valued time series data.

\begin{equation}
\begin{aligned}
\mathcal{D}^c &= \left\{\left(\vx^c_{j}, y^c_{j}\right)\right\}, \ \vx^c_{j} = \left(y^c_{j-a^c}, y^c_{j-a^c +1}, \cdots, y^c_{j-1}\right),\  a^c + 1\leq j\leq T,\\
\mathcal{D}^r &= \left\{\left(\vx^r_{j}, y^r_j\right)\right\}, \ \vx^r_{j} = \left(y^r_{j-a^r}, y^r_{j-a^r +1}, \cdots, y^r_{j-1}\right),\  a^r + 1\leq j\leq T.\\
\end{aligned}
\end{equation}

Based on the constructed datasets \(\mathcal{D}^r\) and \(\mathcal{D}^c\), various supervised learning approaches can be employed for predicting the center time series \(\left\{y^c_{t}\right\}^T_{t=1}\) and the range time series \(\left\{y^r_{t}\right\}^T_{t=1}\) in order to achieve the objective of predicting the interval-valued time series \(\left\{[y^l_{t}, y^u_{t}]\right\}^T_{t=1}\). However, directly modeling the raw data using these models may not yield satisfactory results, as discussed in Section \ref{sec5}. In this paper, we propose a feature-based transfer learning approach to extract features from the data \(\vx^c_j\) (respectively, \(\vx^r_j\)). This approach involves performing a nonlinear transformation on the data \(\vx^c_j\) (respectively, \(\vx^r_j\)) through feature extraction, which enables us to capture the relevant information and patterns that are essential for interval-valued time series prediction. This approach aims to improve the accuracy and performance of interval-valued time series prediction by leveraging the capabilities of transfer learning and feature extraction.

\section{Methodology} \label{sec3}
In this section, we present a comprehensive explanation of our proposed method, which encompasses the construction of the FEN and the feature extraction process using transfer learning.

\subsection{Feature extraction network construction} \label{subsubfea}
In this section, we build a FEN for extracting features from interval-valued time series. The construction of the FEN involves three main steps: data segmentation, imaging, and training.
The first step is to divide the center time series \(\left\{y^c_{t}\right\}^T_{t=1}\) (and similarly for the range time series \(\left\{y^r_{t}\right\}^T_{t=1}\)) into equal parts based on a specified length \(\Delta^c\) (and \(\Delta^r\), respectively). This division process yields a set of segments \(\mathcal{I}^c = \left\{\vw^c_{i}\right\}\) (and \(\mathcal{I}^r = \left\{\vw^r_{i}\right\}\)), where each \(\vw^c_{i}\) (and \(\vw^r_{i}\)) represents a segment of \(\Delta^c\) (and \(\Delta^r\)) length from the center (and range) time series. By dividing the time series data into these segments, we can effectively extract local patterns and dependencies from the data. These segments will serve as inputs to the FEN, which will be responsible for extracting and learning relevant features from the data to be used in the subsequent prediction process. where

\[
\begin{aligned}
\vw^c_{i} &= \left(y^c_{(i-1) \times \Delta^c}, \cdots, y^c_{i \times \Delta^c}\right), 1\leq i\leq \lfloor n/\Delta^c \rfloor, \\
\vw^r_{i} &= \left(y^r_{(i-1) \times \Delta^r}, \cdots, y^r_{i \times \Delta^r}\right), 1\leq i\leq \lfloor n/\Delta^r \rfloor,
\end{aligned}
\]
where \(\lfloor a \rfloor\) denotes the largest integer less than \(a\).

\begin{algorithm}[H]
\caption{Construction of the feature extraction network}
\label{fenc}
\begin{algorithmic}
\REQUIRE ~~\\
Interval-valued time series \(\left\{y_t\right\}^{T}_{t=1}\),\\
Data segmentation length \(\Delta^c\), \(\Delta^r\),\\
Four imaging approaches for time series: RP, GASF, GADF, MTF,\\
Convolutional neural networks: CNNs = \(\left\{\right.\)ResNet18, ResNet34, ResNet50, ResNet101, ResNet152\(\left. \right\}\).\\
\ENSURE ~~\\
Based on CRM \(\longrightarrow \left\{y^c_{t}\right\}^T_{t=1}\) and \(\left\{y^r_{t}\right\}^T_{t=1}\).\\
Based on \(\Delta^c\) and \(\Delta^r\) \( \longrightarrow \mathcal{I}^c\) and \(\mathcal{I}^r\).\\
Based on RP, GASF, GADF and MTF \(\longrightarrow \mathcal{B} = \mathcal{B}^c \cup \mathcal{B}^r\).\\
\textbf{for} item \(=1:\text{length}(\text{CNNs})\)\\
\quad \quad Use CNNs(item) to train \(\mathcal{B}\), \\
\quad \quad Comput training loss \(\mathcal{L}_{\text{item}}\) and test accuracy \(\mathcal{C}_{\text{item}}\).\\
\ \textbf{end}\\
Select the best FEN based on \(\left\{\left(\mathcal{L}_{\text{item}}, \mathcal{C}_{\text{item}}\right)\right\}\), item \(=1, 2, \cdots, \text{length(CNNs)}\).\\
\textbf{Output}: Optimal \(\text{FEN}_\mathcal{B}\).
\end{algorithmic}
\end{algorithm}

After dividing the segments from the center (and range) time series, the next step is to transform these segments into images. Four different imaging approaches can be applied: Recurrence Plot (RP) \cite{eckmann1995recurrence} , Gramian Angular Summation Field (GASF) \cite{wang2015encoding}, Gramian Angular Difference Field (GADF) \cite{wang2015encoding}, and Markov Transition Field (MTF) \cite{wang2015encoding}.  This process results in the construction of an image data set \(\mathcal{A}^c\) (and \(\mathcal{A}^r\)), where each element represents an image obtained from the corresponding segment using one of the imaging approaches. These images capture different visual representations of the underlying patterns and dependencies within the interval-valued time series data. Each imaging approach provides a unique perspective on the data, enabling the extraction of different features and information. The resulting image data sets \(\mathcal{A}^c\) and \(\mathcal{A}^r\) will be utilized in the subsequent steps of the feature extraction procedure.

\[
\begin{aligned}
\mathcal{A}^c &= \left\{\text{RP}(\vw^c_{i}), \text{GASF}(\vw^c_{i}), \text{GADF}(\vw^c_{i}), \text{MTF}(\vw^c_{i})\right\}^{\lfloor n/\Delta^c \rfloor}_{i=1}, \\
\mathcal{A}^r &= \left\{\text{RP}(\vw^r_{i}), \text{GASF}(\vw^r_{i}), \text{GADF}(\vw^r_{i}), \text{MTF}(\vw^r_{i})\right\}^{\lfloor n/\Delta^r \rfloor}_{i=1}.\\
\end{aligned}
\]

Since FEN is an image classification network, it is necessary to assign labels to each image in the dataset \(\mathcal{A}^c\) (respectively, \(\mathcal{A}^r\)) in order to construct an image classification dataset. As each imaging approach captures only a portion of the diverse information contained in the original time series, it is reasonable to assign the same label to the images generated by the same imaging approach. Consequently, we obtain the following dataset, which consists of four classes,
\[
\begin{aligned}
\mathcal{B}^c = &\left\{\left\{\text{RP}(\vw^c_{i}), 1\right\}^{\lfloor n/\Delta^c \rfloor}_{i=1}, \left\{\text{GASF}(\vw^c_{i}), 2\right\}^{\lfloor n/\Delta^c \rfloor}_{i=1}, \right.\\ 
&\left.\left\{\text{GADF}(\vw^c_{i}), 3\right\}^{\lfloor n/\Delta^c \rfloor}_{i=1}, \left\{\text{MTF}(\vw^c_{i}), 4\right\}^{\lfloor n/\Delta^c \rfloor}_{i=1}\right\}, 
\end{aligned}
\]
and \(\mathcal{B}^r\) is the same as above. In general, the time series \(\left\{y^c_{t}\right\}^T_{t=1}\) and \(\left\{y^r_{t}\right\}^T_{t=1}\) should exhibit similar periodic and trend patterns. However, considering the dependency within the interval-valued time series and the requirement for a large amount of data to train the FEN, we merge the datasets \(\mathcal{B}^c\) and \(\mathcal{B}^r\) into a unified dataset \(\mathcal{B} = \mathcal{B}^c \cup \mathcal{B}^r\) to train the FEN. From a representation learning perspective, if a network performs better on the dataset \(\mathcal{B}\), it implies that the network has captured a more effective representation of the data in \(\mathcal{B}\). In this paper, we utilize the training loss and test accuracy as metrics to evaluate the quality of the network's representation of \(\mathcal{B}\). The construction procedure of the FEN is outlined in Algorithm \ref{fenc}.

\begin{figure}[H]
\centering
\includegraphics[scale=0.34]{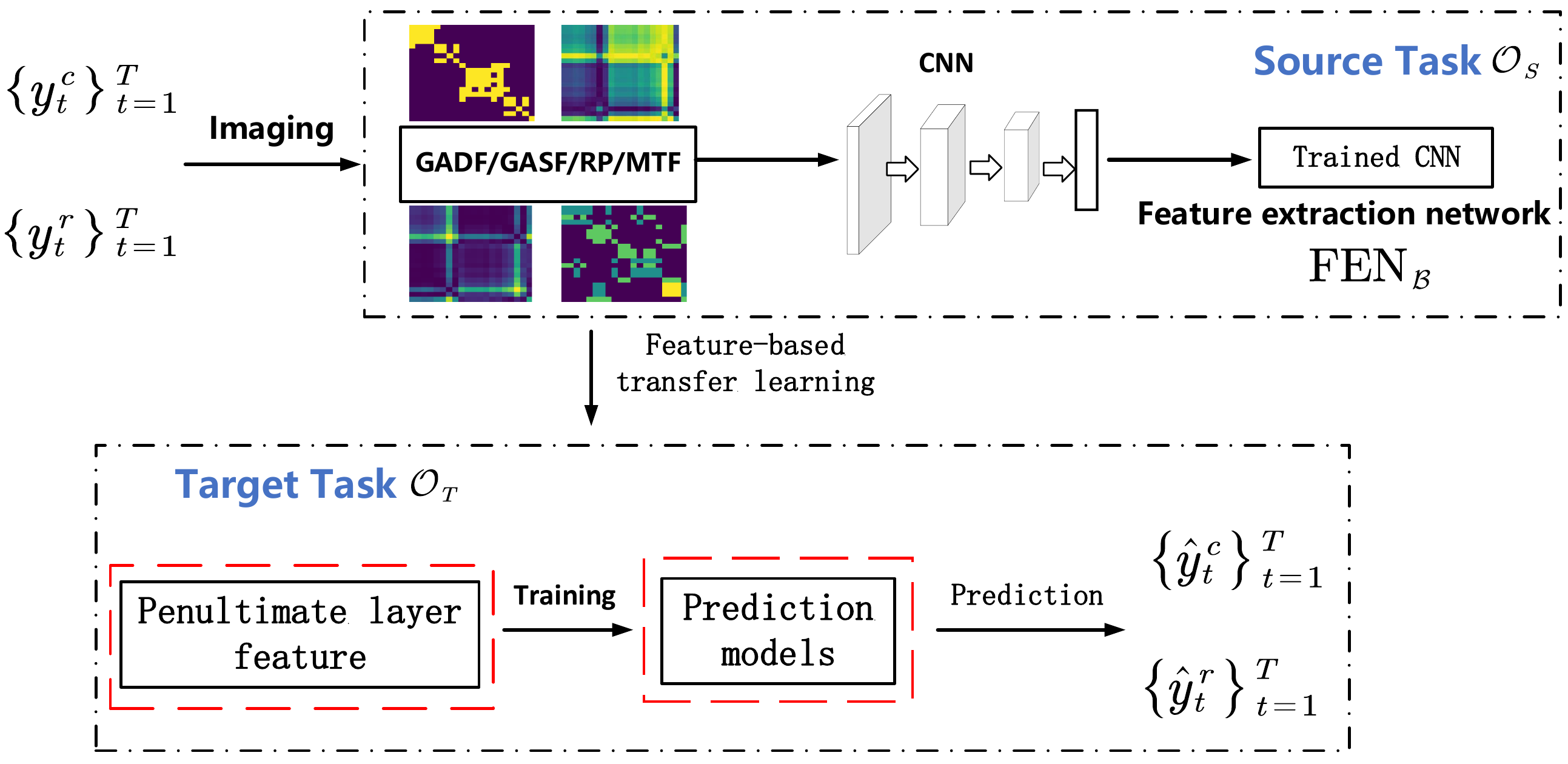}
\caption{\normalsize Feature extraction procedure based on transfer learning. The imaging dataset was initially constructed for the source task \(\mathcal{O}_S\), and the FEN was trained using this dataset. Subsequently, the trained FEN is utilized to extract features, which are then employed for prediction in the target task \(\mathcal{O}_T\). In other words, the features extracted by the FEN from the source task \(\mathcal{O}_S\) are leveraged for prediction in the target task \(\mathcal{O}_T\). } \label{transfer}
\end{figure}

\subsection{Feature extraction based on transfer learning}

Transfer learning involves the process of transferring knowledge from one domain, referred to as the source domain \(\mathcal{O}_S\), to another domain, known as the target domain \(\mathcal{O}_T\). The goal is to utilize the knowledge learned in the source domain \(\mathcal{O}_S\) to enhance the learning performance in the target domain \(\mathcal{O}_T\). In this paper, the source task \(\mathcal{O}_S\) is image classification, while the target task \(\mathcal{O}_T\) is feature extraction. Figure \ref{transfer} illustrates the feature extraction procedure based on feature-based transfer learning. As discussed in Section \ref{subsubfea}, the final feature extraction network \(\text{FEN}_\mathcal{B}\) is determined. When predicting the datasets \(\mathcal{D}^c\) and \(\mathcal{D}^r\), the following procedure is followed. Firstly, the datasets \(\mathcal{D}^c\) and \(\mathcal{D}^r\) are subjected to the imaging process using the four imaging methods. Next, feature extraction is performed based on the feature extraction network \(\text{FEN}_\mathcal{B}\). This step yields the feature-extracted datasets

\[
\begin{aligned}
\mathcal{F}^c = \left\{\left(\text{FEN}_\mathcal{B}\left(\text{Imaging}\left(\vx^c_{j}\right)\right), y^c_{j}\right)\right\}^T_{j = a^c+1},\ \mathcal{F}^r = \left\{\left(\text{FEN}_\mathcal{B}\left(\text{Imaging}\left(\vx^r_{j}\right)\right), y^r_{j}\right)\right\}^T_{j = a^r + 1}.
\end{aligned}
\]

All subsequent training and prediction tasks are carried out using the datasets \(\mathcal{F}^c\) and \(\mathcal{F}^r\). It is important to note that before feature extraction, the data undergoes four different imaging approaches. In Section \ref{sec4}, a comparison of the results obtained using these four imaging approaches is presented. Furthermore, our feature extraction procedure can be combined with any regular regressor for prediction purposes. This means that the extracted features can be utilized as input for a variety of regression models to make predictions in the target task \(\mathcal{O}_T\). The choice of the specific regression model depends on the requirements and characteristics of the target task.

\section{Experiment and Results}\label{sec4}
In this section, we evaluate the proposed feature extraction procedure through Monte Carlo simulations and real data studies. Specifically, we validate the proposed approach using three Data Generating Processes (DGPs) and the S\&P500 index as a real-world dataset.

\subsection{Simulation design} 
In accordance with the model (\ref{datapromodel}), we generate bivariate point-value time series, which represent interval-valued time series in terms of center and range. Following the approach of \textcite{xu2021novel}, we employ the following three settings to generate interval-valued time series: 
\begin{itemize}
\item C1: \(y^c_{t} = 0.4y^c_{(t-1)} + \epsilon^c_t + 2\epsilon^c_{t-1} \),\  \(y^r_{t}\sim \mathcal{U}(30, 50)\).
\item C2:
\[
y^c_{t+2} = \begin{cases}
0.6 + 1.3y^c_{t+1} -0.4y^c_{t} + \epsilon^c_{t+2}, \ y^c_{t} < 5,\\
1.2+1.6y^c_{t+1} - 1.1y^c_{t} + \epsilon^c_{t+2}, \ y^c_{t}\geq 5,
\end{cases}  \  y^c_{t}\sim \mathcal{U}(1, 5).
\] 
\item C3: \(y^r_{t} = 0.2 y^r_{t-1} + 1.6\log (1000y^r_{t-1})+ 30+\epsilon^r_t\),\   \(y^r_{t}\sim \mathcal{U}(20, 70)\).
\end{itemize}

\begin{figure}[H]
\centering 
\begin{minipage}[t]{0.48\textwidth}
\centering
\includegraphics[scale=0.89]{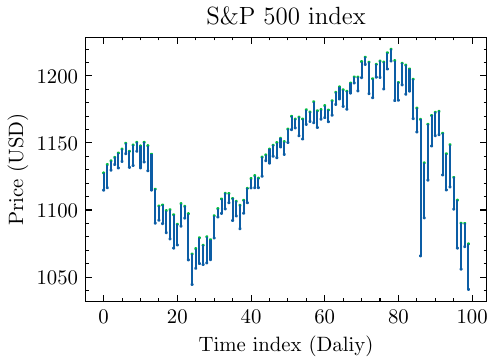}
\end{minipage}
\begin{minipage}[t]{0.48\textwidth}
\centering
\includegraphics[scale=0.89]{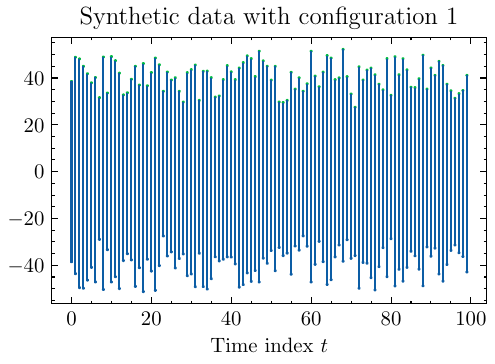}
\end{minipage}
\begin{minipage}[t]{0.48\textwidth}
\centering
\includegraphics[scale=0.89]{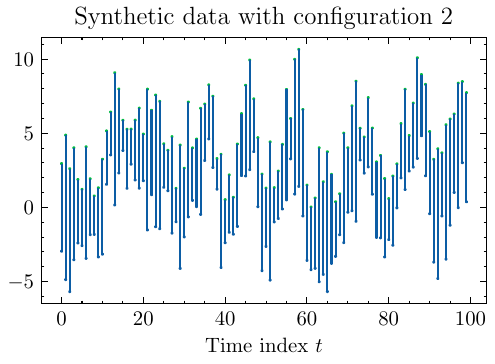}
\end{minipage}
\begin{minipage}[t]{0.48\textwidth}
\centering
\includegraphics[scale=0.89]{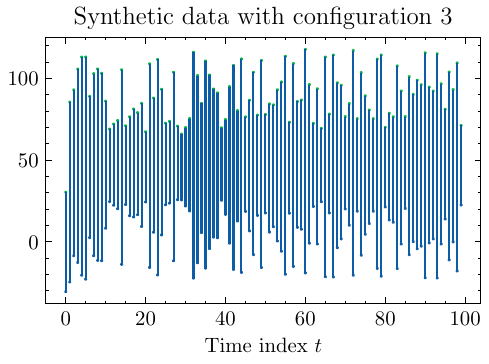}
\end{minipage}
\caption{\normalsize The first 100 observation time series graphs of S\&P500 index and three DGPs.}\label{spthreesy}	
\end{figure}

In C1, C2, and C3, \(\epsilon_t\) represents independent and identically distributed random variables following the standard normal distribution \(\mathcal{N}(0,1)\). Additionally, \(\mathcal{U}(a,b)(a<b)\) denotes the uniform distribution between \(a\) and \(b\). The length of the interval-valued time series generated under all three settings is 1500. For C1, when \(t=1\), the iteration starts with \(y_0=0\). For C2, the iteration starts with \(y_0=y_1=0\). Lastly, for C3, the iteration starts with \(y_0=0.001\). Regarding the real data, we have selected the daily data of the S\&P500\footnote{Historical data on the S\&P500 index can be obtained from the website \url{https://finance.yahoo.com/}.} index from January 1, 2012, to August 27, 2021. The S\&P500 index data consists of daily open and close prices, as well as high and low prices. In this study, the high and low prices are chosen for analysis and prediction. Figure \ref{spthreesy} displays the time series graphs of the first 100 observations of the S\&P500 index and the three DGPs.
%The following provides an illustration and visualization of the DGPs and the S\&P500 index\footnote{Historical data on the S\&P500 index can be obtained from the website \url{https://finance.yahoo.com/}.}.

In the following numerical experiments, we employ several evaluation metrics to assess the performance of the models and the proposed feature extraction procedure. The selected metrics include Mean Square Error (MSE), Mean Absolute Error (MAE), Symmetric Mean Absolute Percentage Error (SMAPE) and Mean Distance Error (MDE), which are defined as
\[
\begin{aligned}
\text{MSE} &= \frac{1}{T} \sum_{t=1}^{T}\left(y^c_{t} - \widehat{y}^c_{t}\right)^2, \ \text{MAE} = \frac{1}{T}\sum_{t=1}^{T} \left|y^c_{t} - \widehat{y}^c_{t}\right|,\\
\text{SMAPE} &= \frac{1}{T}\sum_{t=1}^{T}\frac{\left|y^c_{t} - \widehat{y}^c_{t}\right|}{|y^c_{t}| + |\widehat{y}^c_{t}|}, \ \text{MAPE} = \frac{1}{T} \sum_{t=1}^{T} \left| \frac{y^c_{t} - \widehat{y}^c_{t}}{y^c_{t}}\right|,
\end{aligned}
\]

\[
\text{MDE} = \sqrt{\frac{1}{T}\sum_{t=1}^{T}\sqrt{\frac{\left((y^c_{t} - y^r_{t}) - (\widehat{y}^c_{t} - \widehat{y}^r_{t})\right)^2 + \left((y^r_{t} + y^c_{t}) - (\widehat{y}^r_{t} + \widehat{y}^c_{t})\right)^2}{2}}},
\]
where \(\left\{\widehat{y}^c_{t}\right\}^T_{t=1}\) and \(\left\{\widehat{y}^r_{t}\right\}^T_{t=1}\) represent the predicted values corresponding to the center \(\left\{y^c_{t}\right\}^T_{t=1}\) and range \(\left\{y^r_{t}\right\}^T_{t=1}\), respectively. To verify the effectiveness of the proposed feature extraction procedure \(\text{FEN}(\cdot)\) and demonstrate its superiority over general deep learning models, we considered three different approaches for predicting the raw data.

The first approach involves making predictions directly on the original data without any preprocessing or feature extraction steps. In this approach,  we employ several common statistical learning models to directly perform predictions on the raw data sets. The chosen methods include Support Vector Machine (SVM) \cite{SmolaAlextutorialsvr2004}, Random Forest (RF) \cite{CutlerAdele2011}, XGBoost \cite{DBLP:conf/kdd/ChenG16},  Wavelet Neural Network (WNN) \cite{AlexandridisAntonios2011} and Long Short-Term Memory Network (LSTM) \cite{HochreiterSepp1997}. SVM is a popular machine learning algorithm that can be used for classification or regression tasks. It aims to find a hyperplane that maximally separates the data points. RF is an ensemble learning method that constructs multiple decision trees and combines their predictions to make accurate predictions. It is known for its robustness and ability to handle high-dimensional data. XGBoost is a gradient boosting framework that is widely used for classification and regression tasks. It leverages the strengths of gradient boosting algorithms and employs a more efficient optimization algorithm. WNN combines wavelet analysis and neural networks to capture both local and global features in the data. It has been successfully applied in various time series prediction tasks. LSTM is a type of recurrent neural network (RNN) that is capable of learning long-term dependencies in sequential data. It has been widely used for time series prediction tasks due to its ability to capture temporal patterns. For more detailed information, please refer to Subsection \ref{rawset} for the experimental setup.

The second approach involves combining the aforementioned models with the feature extraction procedure \(\text{FEN}(\cdot)\) for predicting the raw data sets. The specific details of this combination can be found in Subsections \ref{fencon} and \ref{fenp}. By incorporating the feature extraction procedure \(\text{FEN}(\cdot)\) into the prediction process of the selected models, we aim to demonstrate the enhanced performance and effectiveness of this combined approach for predicting the raw data sets.

The third method involves selecting several representative deep learning-based approaches for time series prediction on the raw data sets. These approaches include: N-BEATS \cite{DBLP:conf/iclr/OreshkinCCB20}: N-BEATS is a deep learning architecture specifically designed for time series forecasting. It utilizes fully connected neural networks and a stack of blocks to capture temporal dependencies in the data. N-HiTS \cite{DBLP:journals/corr/abs-2201-12886}: N-HiTS is a hybrid deep learning model that combines both convolutional and recurrent neural networks. It leverages the strengths of both architectures to capture local and global features in the time series data. Temporal Convolutional Network (TCN) \cite{DBLP:journals/corr/abs-1803-01271}: TCN is a deep learning architecture that employs dilated convolutions to process sequential data. It can capture long-term dependencies in the time series while being computationally efficient. Transformer \cite{DBLP:conf/nips/VaswaniSPUJGKP17}: Transformer is a state-of-the-art deep learning model originally designed for natural language processing tasks. It utilizes self-attention mechanisms to capture global dependencies in the input sequence and has been successfully applied to time series forecasting. More detailed information regarding these deep learning-based approaches and their integration with the raw data sets can be found in Subsection \ref{otherdeep}.

\begin{figure}[H]
\centering 
% S&P500 index
\begin{minipage}[t]{0.24\textwidth}
\centering
\includegraphics[scale=0.525]{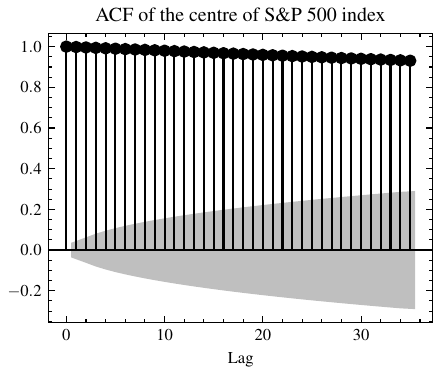}
\end{minipage}
\begin{minipage}[t]{0.24\textwidth}
\centering
\includegraphics[scale=0.525]{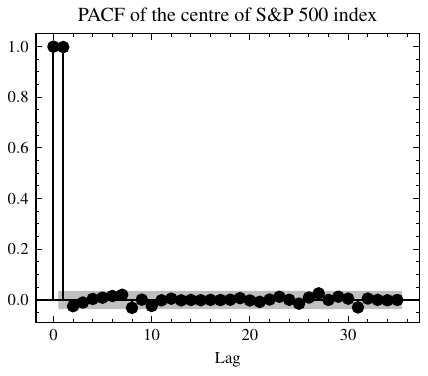}
\end{minipage}
\begin{minipage}[t]{0.24\textwidth}
\centering
\includegraphics[scale=0.525]{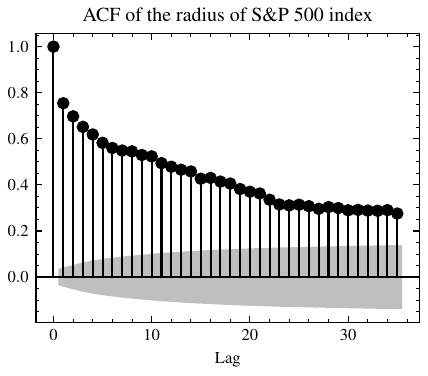}
\end{minipage}
\begin{minipage}[t]{0.24\textwidth}
\centering
\includegraphics[scale=0.525]{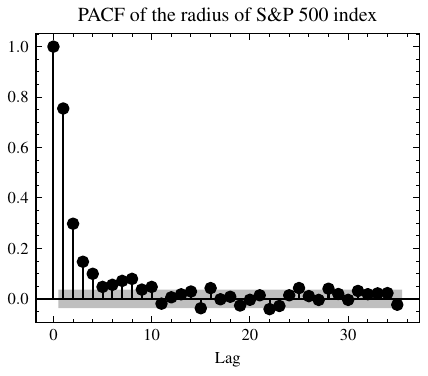}
\end{minipage}

% C1
\begin{minipage}[t]{0.24\textwidth}
\centering
\includegraphics[scale=0.525]{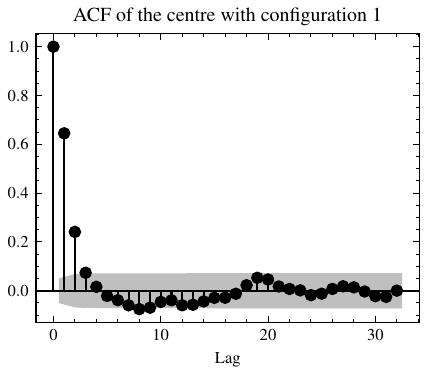}
\end{minipage}
\begin{minipage}[t]{0.24\textwidth}
\centering
\includegraphics[scale=0.525]{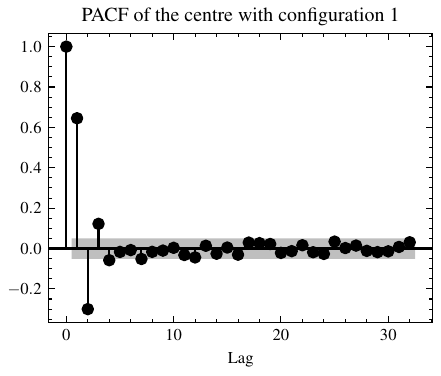}
\end{minipage}
\begin{minipage}[t]{0.24\textwidth}
\centering
\includegraphics[scale=0.525]{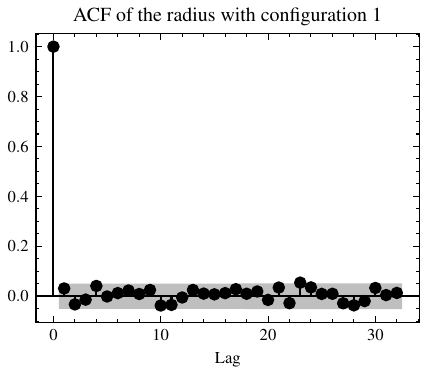}
\end{minipage}
\begin{minipage}[t]{0.24\textwidth}
\centering
\includegraphics[scale=0.525]{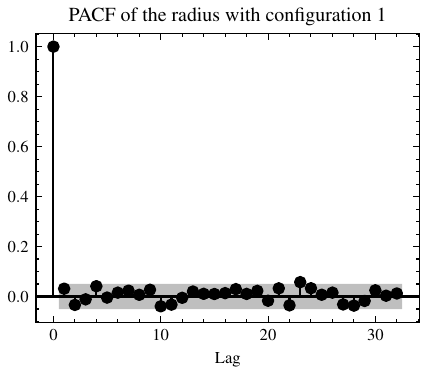}
\end{minipage}

% C2
\begin{minipage}[t]{0.24\textwidth}
\centering
\includegraphics[scale=0.525]{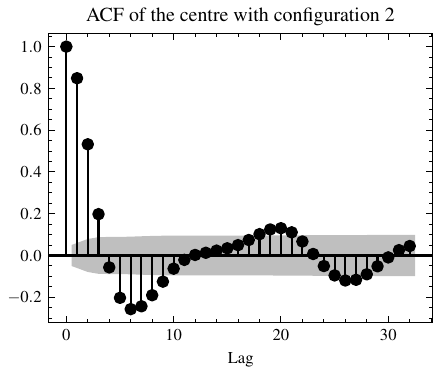}
\end{minipage}
\begin{minipage}[t]{0.24\textwidth}
\centering
\includegraphics[scale=0.525]{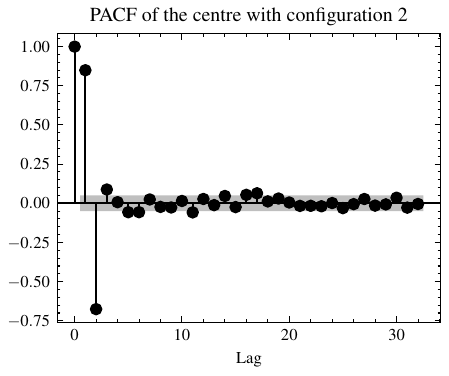}
\end{minipage}
\begin{minipage}[t]{0.24\textwidth}
\centering
\includegraphics[scale=0.525]{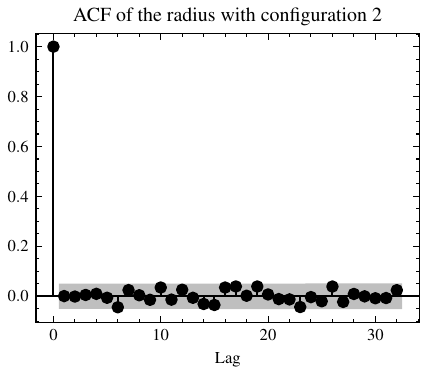}
\end{minipage}
\begin{minipage}[t]{0.24\textwidth}
\centering
\includegraphics[scale=0.525]{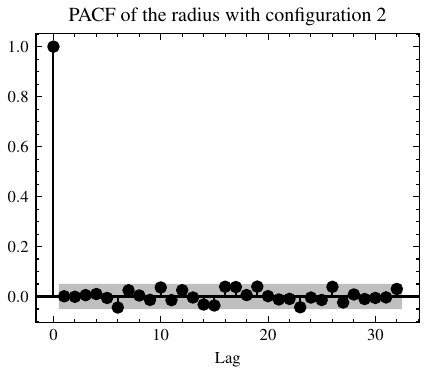}
\end{minipage}

% C3
\begin{minipage}[t]{0.24\textwidth}
\centering
\includegraphics[scale=0.525]{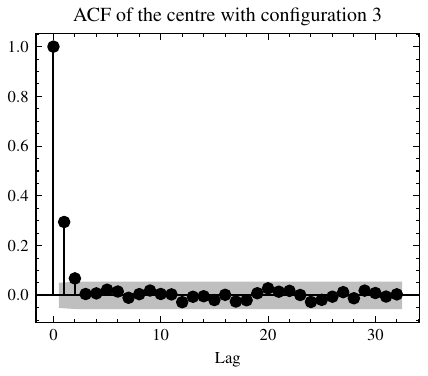}
\end{minipage}
\begin{minipage}[t]{0.24\textwidth}
\centering
\includegraphics[scale=0.525]{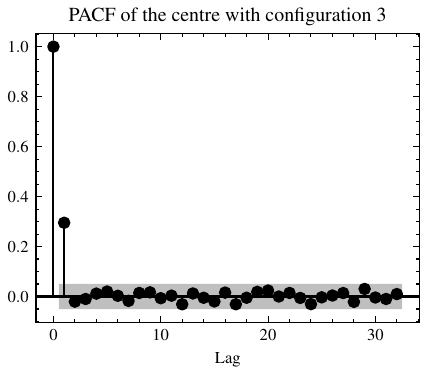}
\end{minipage}
\begin{minipage}[t]{0.24\textwidth}
\centering
\includegraphics[scale=0.525]{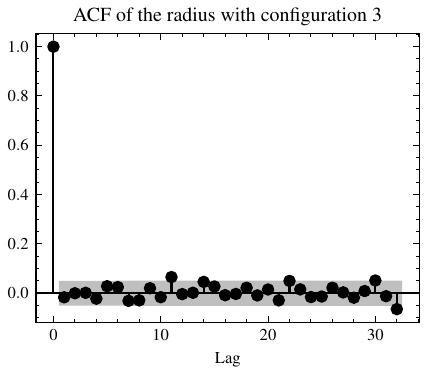}
\end{minipage}
\begin{minipage}[t]{0.24\textwidth}
\centering
\includegraphics[scale=0.525]{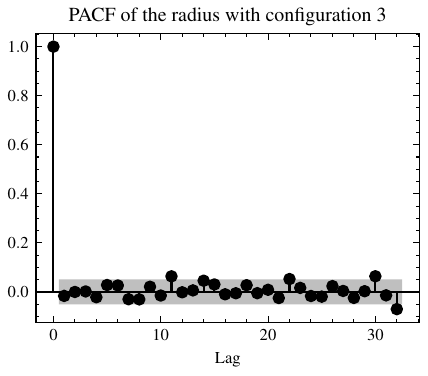}
\end{minipage}
\caption{\normalsize The ACF and PACF of the centers and ranges of the samples corresponding to the S\&P500 index and three DGPs. Each row represents a specific dataset or DGP. }\label{ACFPACF}	
\end{figure} 

\subsection{Raw datasets prediction results} \label{rawset}

As depicted in the technology roadmap \ref{technicalroadmap}, it is necessary to determine the appropriate order before performing prediction. In this paper, two model-free order determination measures, namely the ACF and PACF, are chosen for this purpose. Figure \ref{ACFPACF} illustrates the ACF and PACF of the centers and ranges of the samples corresponding to the S\&P500 index and the three DGPs. Based on these plots, the orders corresponding to the time series are determined, as shown in Table \ref{ACFPACFording}.
\begin{table}[H]
\centering
\caption{\normalsize The appropriate orders for the center and range corresponding to the S\&P500 index and the three DGPs are determined based on the analysis of the ACF and PACF plots.} \label{ACFPACFording}
\begin{tabular}{ccccccccc} 
\toprule
Datasets & S\&P500-\(r\)&S\&P500-\(c\)& C1-\(c\)& C1-\(r\) & C2-\(c\)& C2-\(r\) & C3-\(c\)& C3-\(r\)\\
\midrule
Order & 35& 35 & 5&3 & 25& 5& 5&5 \\
\bottomrule 
\end{tabular}
\begin{tablenotes}
\item[1] -\(c\) and -\(r\) denote centre and range respectively.
\end{tablenotes}
\end{table}

The prediction results presented in Table \ref{originresults} are based on the order determination specified in Table \ref{ACFPACFording} and the five models mentioned earlier. These results pertain to the S\&P500 index and the three DGPs under consideration.

\begin{table}[H]
\footnotesize
\centering
\caption{\normalsize Raw datasets prediction results}
\label{originresults} 
\setlength{\tabcolsep}{1.3mm}{  
\begin{tabular}{clccccccccc}
\toprule
\multirow{2}{*}{\tabincell{c}{Datasets}}&\multirow{2}{*}{Models}&\multicolumn{2}{c}{MSE}  & \multicolumn{2}{c}{MAE} & \multicolumn{2}{c}{MAPE} 
& \multicolumn{2}{c}{SMAPE} & MDE\\
\cline{3-11}
& & center & range & center & range & center & range & center & range & interval\\ 
% data set C1
\midrule
\multirow{5}{*}{C1} & SVR &4.5988&34.0869&1.6902&5.0614&1.7602&0.1291&1.1007&0.1273&2.3705\\
& RF &4.8064&33.6207&1.7233&\textbf{1.0582}&2.1668&0.1287&1.0582&0.1266&2.3696\\
& LSTM &\textbf{4.2308}&\textbf{33.5927}&\textbf{1.6183}&5.0193&1.8027&\textbf{0.1285}&\textbf{1.0579}&\textbf{0.1263}&\textbf{2.3575}\\
& XGBoost &7.6799&33.6633&2.1733&5.0486&\textbf{1.0280}&0.1291&1.6686&0.1270&2.4270\\
& WNN &4.3078&33.6082&1.6323&5.0421&2.7989&0.1292&1.4250&0.1268&47.8792\\
\midrule
% data set C2
\multirow{5}{*}{C2} & SVR &4.9823&\textbf{35.1249}&1.7574&5.1206&\textbf{1.6823}&0.1305&1.1556&0.1288&2.3897\\
& RF &4.8320&33.9315&1.7286&5.0646&2.1768&0.1296&1.0465&0.1274&2.3756\\
& LSTM &\textbf{4.0808}&33.8990&\textbf{1.6098}&\textbf{5.0493}&1.9107&0.1293&\textbf{1.0379}&\textbf{0.1270}&\textbf{2.3592}\\
& XGBoost &6.0233&122.9581&1.9315&9.4900&2.4109&0.2200&1.1684&0.2563&3.1584\\
& WNN &8.6832&33.6960&2.3090&5.0518&2.0419&\textbf{0.1271}&1.4985&0.1271&47.1844\\
\midrule
% data set C3
\multirow{5}{*}{C3} & SVR &4.5988&35.1249&1.6902&5.1206&1.7602&0.1305&1.1007&0.1288&2.3822\\
& RF &4.8064&33.9315&1.7233&5.0646&2.1668&0.1296&\textbf{1.0582}&0.1274&\textbf{2.3744}\\
& LSTM &\textbf{4.1607}&35.4555&\textbf{1.6129}&5.1708&1.8393&0.1303&1.0731&0.1301&2.3828\\
& XGBoost &7.6799&33.7487&2.1733&5.0579&\textbf{1.0280}&0.1293&1.6686&0.1272&2.4271\\
& WNN &7.5627&\textbf{33.6960}&2.1760&\textbf{5.0518}&1.5784&\textbf{0.1271}&1.5295&\textbf{0.1271}&47.1020\\
\midrule
% data set SP500
\multirow{5}{*}{S\&P500} & SVR &4.9309&34.8637&1.7406&5.0842&\textbf{1.6122}&0.1289&\textbf{1.1509}&0.1277&2.3843\\
& RF &4.8074&33.9036&\textbf{1.7198}&5.0737&2.0349&0.1292& 1.0560&0.1274&\textbf{2.3759}\\
& LSTM &\textbf{4.1061}&558.8109&1.6023&20.0523&1.6898&0.5104&1.0519&0.5683&4.5013\\
& XGBoost &4.9526&47.0330&1.7298&5.8277&1.9688&0.1473&1.0832&0.1472&2.5275\\
& WNN &9.8931&\textbf{33.5523}&2.5065&\textbf{5.0392}&2.3020&\textbf{0.1286}&1.4750&\textbf{0.1266}&47.1248\\
\bottomrule
\end{tabular}
}
\end{table}	

The table above clearly demonstrates that the LSTM model outperforms the other four models in terms of predicting the raw datasets. In particular, for the center and range time series of DGP1, the LSTM exhibits superior performance across almost all evaluation metrics, closely followed by the WNN model. The strong predictive capability of the LSTM and WNN can be attributed to their recursive structures, which enable them to effectively capture the dependency patterns inherent in the DGPs. It is important to note that while the LSTM and WNN models demonstrate better prediction performance, the performance of all five models on the four raw datasets is relatively comparable. With parameter tuning and optimization, it may be possible to narrow the performance gap among the models. Therefore, further refinement and exploration of the models' parameters could potentially lead to improved prediction results. Overall, the results indicate that the LSTM model, followed by the WNN model, exhibits the most promising performance for predicting the raw data sets. 

\subsection{Feature extraction network construction} \label{fencon}

\begin{figure}[H]
\centering
\includegraphics[scale=0.98]{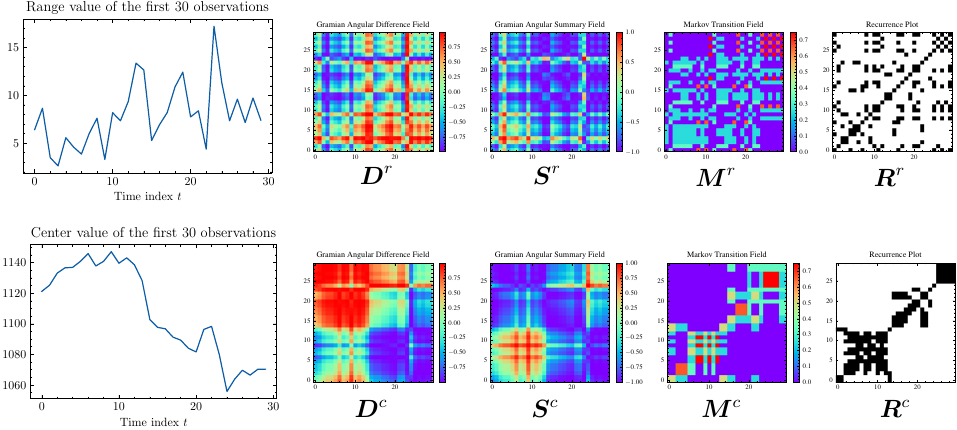}
\caption{\normalsize A visual demonstration of the four imaging methods on the first 30 observations of S\&P500 index. The first column of the figure displays the time-series graph of the range \(\left\{y^r_{t}\right\}^{30}_{t=1}\) and center \(\left\{y^c_{t}\right\}^{30}_{t=1}\). The subsequent four columns depict the images obtained using the GADF, GASF, MTF, and RP methods, respectively.} \label{imagemeth}
\end{figure}

In this subsection, we focus on a crucial aspect of the proposed method, which is the construction of the FEN. Following the approach of \textcite{tian2021drought}, we consider six specific lengths (30, 35, 40, 45, 50, and 55) to disjointly segment the central time series and the range time series. Subsequently, an imaging dataset is generated based on the four imaging approaches. An example of the four imaging methods applied to the S\&P500 index is depicted in Figure \ref{imagemeth}.

To build the FEN, we select different depths of the deep residual network \cite{he2016deep}, namely ResNet18, ResNet34, ResNet50, ResNet101, and ResNet152, as the backbone architectures. Each DGP requires only one FEN, and hence we determine the appropriate FEN and segmentation length for each DGP. To achieve optimal performance, we evaluate the three metrics of training loss, test accuracy, and epoch. The results are summarized in Tables \ref{fenselect1} and \ref{fenselect2}, which provide insights into the optimal FEN and segmentation length for each DGP. By considering these metrics, we ensure that the selected FEN and segmentation length yield the best performance for each DGP. This optimization process enables us to construct an efficient and effective FEN tailored to the specific characteristics of each DGP, enhancing the overall predictive capabilities of the proposed method.

\begin{table}[H]
\footnotesize
\centering
\caption{\normalsize The training results of ResNet models with different depths (ResNet18, ResNet34, ResNet50, ResNet101, and ResNet152) on C1 and C2 datasets under different segmentation lengths (All numbers in the tables are expressed as percentages except Best Epoch. The best results obtained for each metric are bolded.)}\label{fenselect1} 
\begin{tabular}{clcc}
\toprule
\multirow{3}{*}{\tabincell{c}{CNNs}}&\multirow{3}{*}{Metrics}&\multicolumn{2}{c}{Training results}\\
&&\multicolumn{2}{c}{C1 \ \ \ \ \ C2}\\
&&\multicolumn{2}{c}{30/35/40/45/50/55}\\
\midrule
\multirow{3}{*}{\tabincell{c}{ResNet-18}}&Training Loss& 0.845/0.837/1.138/\textbf{0.608}/3.163/1.529     &0.492/0.813/1.986/1.086/0.496/0.868  \\	
&Test Accuracy & 0.760/1.000/0.946/\textbf{1.000}/0.967/0.630  &    0.910/0.750/0.784/0.939/0.967/0.759 \\
&Best Epoch &10/10/7/\textbf{10}/4/9   &      8/9/9/8/9/8  \\

\midrule
\multirow{3}{*}{\tabincell{c}{ResNet-34}}&Training Loss& 0.380/0.337/0.414/0.902/0.340/1.488   &    0.671/1.042/0.795/\textbf{0.311}/0.330/0.954 \\
&Test Accuracy &1.000/1.000/1.000/0.758/1.000/0.704  & 0.950/0.821/1.000/\textbf{1.000}/1.000/0.981    \\
&Best Epoch & 7/7/8/10/10/6    &9/7/8/\textbf{8}/10/10    \\

\midrule
\multirow{3}{*}{\tabincell{c}{ResNet-50}}&Training Loss&0.176/0.247/0.183/0.329/0.159/0.698  &   0.171/0.175/0.296/0.747/0.219/0.801    \\
&Test Accuracy &1.000/1.000/1.000/1.000/1.000/1.000  &  0.990/1.000/1.000/0.955/1.000/0.944 \\
&Best Epoch &7/5/8/10/10/8   & 10/6/10/9/10/6    \\

\midrule
\multirow{3}{*}{\tabincell{c}{ResNet-101}}&Training Loss&0.255/1.000/0.884/0.775/0.153/0.561   & 0.216/0.294/0.231/0.716/0.433/0.903         \\
&Test Accuracy &1.000/0.98/0.946/1.000/1.000/0.778  &  0.990/1.000/1.000/1.000/1.000/0.963	  \\
&Best Epoch &7/10/10/4/10/10   &  9/10 /8/8/10/5    \\

\midrule
\multirow{3}{*}{\tabincell{c}{ResNet-152}}&Training Loss&0.515/0.339/0.407/0.244/0.465/0.561  &   0.774/0.912/0.716/0.225/0.656/0.716   \\
&Test Accuracy &0.980/1.000/1.000/1.000/0.967/0.778   & 0.890/0.869/0.784/0.985/0.767/0.784  \\
&Best Epoch &6/10/10/10/8/6  &  7/9/8/10/9/8  \\
\bottomrule
\end{tabular}
\end{table}

\begin{table}[H]
\footnotesize
\centering
\caption{\normalsize The training results of ResNet models with different depths (ResNet18, ResNet34, ResNet50, ResNet101, and ResNet152) on C3 and S\&P500 datasets under different segmentation lengths (All numbers in the tables are expressed as percentages except Best Epoch. The best results obtained for each metric are bolded.)}\label{fenselect2}
\begin{tabular}{clcc}
\toprule
\multirow{3}{*}{\tabincell{c}{CNNs}}&\multirow{3}{*}{Metrics}&\multicolumn{2}{c}{Training results}\\
&&\multicolumn{2}{c}{C3 \ \ \ \ \ S\&P500}\\
&&\multicolumn{2}{c}{30/35/40/45/50/55}\\
\midrule
\multirow{3}{*}{\tabincell{c}{ResNet-18}}&Training Loss&  0.842/1.009/0.778/2.342/2.447/0.946   &   0.582/0.590/0.603/1.121/0.586/0.734 \\	
&Test Accuracy &  0.880/0.821/0.986/0.955/1.000/0.593  &0.948/0.910/0.952/0.969/0.940/0.887 \\
&Best Epoch &   9/8/9/10/10/10   & 5/9/10/3/9/9  \\

\midrule
\multirow{3}{*}{\tabincell{c}{ResNet-34}}&Training Loss&      0.746/0.393/1.139 /0.267/0.510/2.447 &  0.974/0.717/0.518/0.552/0.583/0.596  \\
&Test Accuracy &  0.790/1.000/1.000/1.000/1.000/0.685 &  0.964/0.892/0.973/0.962/0.966/0.915   \\
&Best Epoch &   7/9/9/4/6/4   &  3/4/10/9/10/10 \\

\midrule
\multirow{3}{*}{\tabincell{c}{ResNet-50}}&Training Loss&  0.415/\textbf{0.126}/0.198/0.311/0.395/1.016  &  0.343/0.564/\textbf{0.506}/0.455/0.576/0.902\\
&Test Accuracy  &  1.000/\textbf{1.000}/1.000/1.000/1.000/0.963  &  0.954/0.964/\textbf{0.979}/0.969/0.914/0.915 \\
&Best Epoch &    10/\textbf{10}/10/4/10/9  &  10/5/\textbf{8}/7/5/4 \\

\midrule
\multirow{3}{*}{\tabincell{c}{ResNet-101}} &Training Loss &    0.471/0.235/0.530/0.265/0.413/0.958     &  1.121/0.776/0.369/0.471/0.531/0.736 \\
&Test Accuracy &   1.000/1.000/1.000/1.000/1.000/0.926      &  0.969/0.705/0.966/0.954/0.888 /0.906  \\
&Best Epoch &    9/9/7/10/10/8   &   2/4/10/10/8/5\\

\midrule
\multirow{3}{*}{\tabincell{c}{ResNet-152}}&Training Loss &   0.252/0.256/0.174/0.504/0.087/0.724   &   0.422/0.582/0.533/0.428/0.646/0.530 \\
&Test Accuracy  &  1.000/1.000/1.000/0.758/1.000/0.685    &  0.959/0.970/0.959/0.969/0.922/0.934  \\
&Best Epoch  &   6/9/7/10/9/10  &  8/6/7/8/9/9 \\
\bottomrule
\end{tabular}
\end{table}

From Tables \ref{fenselect1} and \ref{fenselect2}, we identify the optimal combination of network structure and segmentation length for each DGP based on the training loss, test accuracy, and optimal training epoch. We also prioritize a shallow network and a relatively small training epoch, provided it achieves comparable classification performance. The ultimate combination of FEN structure, optimal training epoch, and data segmentation length for each DGP is determined as follows:
\begin{itemize}
\item C1: ResNet-18, 10, 45; C2: ResNet-34, 8, 45; C3: ResNet-50, 10, 35;	
\item S\&P500: ResNet-50, 8, 40.
\end{itemize}
Other hyperparameters, such as learning rate, optimization algorithm, and batch size, remain consistent during the training of ResNet models with different depths for different imaging datasets. By keeping these hyperparameters constant, we ensure a fair and unbiased comparison between the ResNet models.

\subsection{Prediction results based on FEN} \label{fenp}

In Subsection \ref{fencon}, we have determined the final FENs for the four DGPs. In this subsection, we combine the feature extraction procedure with the models used in Subsection \ref{rawset} to predict the four DGPs. The FENs for the four DGPs are denoted as FEN\(_1\), FEN\(_2\), FEN\(_3\), and FEN\(_4\), and the corresponding segmentation lengths are \(\ell_1\), \(\ell_2\), \(\ell_3\), and \(\ell_4\). Given the determination of the four imaging approaches, the imaging data set for training the FENs is determined by the segmentation length. To preserve the information without loss during the feature extraction phase, we specify the order of the DGP as the segmentation length. Consequently, we can construct the following data sets based on the trained FENs:

\[
\begin{aligned}
\mathcal{F}^c_i &= \left\{\left(\text{FEN}_i\left(\text{Imaging}\left(\vx^c_{j}\right)\right), y^c_{j}\right)\right\}_{j = \ell_i + 1}^T,\\
\mathcal{F}^r_i &= \left\{\left(\text{FEN}_i\left(\text{Imaging}\left(\vx^r_{j}\right)\right), y^r_{j}\right)\right\}_{j = \ell_i+1}^T, 
\end{aligned}
\]
where \(i = 1,2,3,4\). Based on the FEN and the four imaging approaches, a time series can be represented in four different ways. The prediction performance of the five models on each of these four representations is presented in Tables \ref{fenresulsR}, \ref{fenresulsS}, \ref{fenresulsD}, and \ref{fenresulsM}. 
\begin{table}[H]
\footnotesize
\centering
\setlength{\abovecaptionskip}{0pt}%
\setlength{\belowcaptionskip}{3pt}%
\caption{\normalsize The prediction performance of the models on features extracted from the RP representation}\label{fenresulsR}
\begin{tabular}{clccccccccc}
\toprule
\multirow{2}{*}{\tabincell{c}{Data sets\\ (FEN)}}&\multirow{2}{*}{Models}&\multicolumn{2}{c}{MSE}  & \multicolumn{2}{c}{MAE} & \multicolumn{2}{c}{MAPE} 
& \multicolumn{2}{c}{SMAPE} & MDE\\
\cline{3-11}
& & center & range & center & range & center & range & center & range & interval\\
% data set C1
\midrule
\multirow{5}{*}{\tabincell{c}{C1 \\ (ResNet-18)} } & SVR &0.0211 & 0.0897 & 0.1155 & 0.2605 & 0.2233 & 2.2243 & 0.2087 & 0.5940 & 0.5526\\
& RF &0.0198 & 0.0852 & 0.1108 & 0.2547 & 0.2180 & 2.9678 & 0.2001 & 0.5819 & 0.5451\\
& LSTM &0.0199 & 0.0853 & 0.1114 & 0.2541 & 0.2200 & 2.4456 & 0.2010 & 0.5803 & 0.5451\\
& XGBoost &0.0221 & 0.1005 & 0.1179 & 0.2718 & 0.2310 & 2.5893 & 0.2129 & 0.6147 & 0.5623\\
& WNN &0.0199 & 0.0850 & 0.1107 & 0.2544 & 0.2177 & 1.9891 & 0.1995 & 0.5805 & 10.3897\\
\midrule
% data set C2
\multirow{5}{*}{\tabincell{c}{C2\\ (ResNet-34)} } & SVR &0.0291 & 0.0887 & 0.1382 & 0.2585 & 0.3955 & 3.1962 & 0.2844 & 0.5985 & 0.5588\\
& RF &0.0287 & 0.0848 & 0.1365 & 0.2551 & 0.4063 & 3.0555 & 0.2804 & 0.5948 & 0.5551\\
& LSTM &0.0291 & 0.0877 & 0.1376 & 0.2585 & 0.4042 & 3.0716 & 0.2826 & 0.5981 & 0.5586\\
& XGBoost &0.0341 & 0.1023 & 0.1461 & 0.2695 & 0.4233 & 3.0549 & 0.3002 & 0.6232 & 0.5757\\
& WNN &0.0289 & 0.0871 & 0.1368 & 0.2588 & 0.4073 & 3.0675 & 0.2808 & 0.5995 & 10.6483\\
\midrule
% data set C3
\multirow{5}{*}{\tabincell{c}{C3\\ (ResNet-50)}} & SVR &0.0267 & 0.0902 & 0.1311 & 0.2606 & 0.4544 & 3.3431 & 0.2752 & 0.5990 & 0.5593\\
& RF &0.0255 & 0.0863 & 0.1288 & 0.2578 & 0.4677 & 2.6781 & 0.2707 & 0.5927 & 0.5559\\
& LSTM &0.0260 & 0.0866 & 0.1293 & 0.2589 & 0.7763 & 3.8982 & 0.2722 & 0.5958 & 0.5567\\
& XGBoost &0.0276 & 0.1043 & 0.1332 & 0.2741 & 0.2231 & 3.2212 & 0.2790 & 0.6210 & 0.5722\\
& WNN &0.0257 & 0.0858 & 0.1295 & 0.2571 & 0.3321 & 3.5477 & 0.2728 & 0.5925 & 10.6564\\
\midrule
% data set SP500
\multirow{5}{*}{\tabincell{c}{S\&P500 \\ (ResNet-50)}} & SVR &0.1781 & 0.0194 & 0.3914 & 0.1080 & 0.5812 & 1.2767 & 0.8336 & 0.6330 & 0.6464\\
& RF &0.1931 & 0.0274 & 0.4109 & 0.1056 & 0.6130 & 0.5435 & 0.8954 & 0.7073 & 0.6658\\
& LSTM &0.1926 & 0.0270 & 0.4104 & 0.1042 & 0.6125 & 0.5416 & 0.8941 & 0.6920 & 0.6651\\
& XGBoost &0.2019 & 0.0281 & 0.4142 & 0.1079 & 0.6167 & 0.5678 & 0.9236 & 0.7488 & 0.6691\\
& WNN &0.1926 & 0.0276 & 0.4105 & 0.1060 & 0.6124 & 0.5444 & 0.8939 & 0.7109 & 17.9086\\
\bottomrule
\end{tabular}
\end{table}

\begin{table}[H]
\footnotesize
\centering
\setlength{\abovecaptionskip}{0pt}%
\setlength{\belowcaptionskip}{3pt}%
\caption{\normalsize The prediction performance of the models on features extracted from the GASF representation}\label{fenresulsS}
\begin{tabular}{clccccccccc}
\toprule
\multirow{2}{*}{\tabincell{c}{Data sets\\ (FEN)}}&\multirow{2}{*}{Models}&\multicolumn{2}{c}{MSE}  & \multicolumn{2}{c}{MAE} & \multicolumn{2}{c}{MAPE} 
& \multicolumn{2}{c}{SMAPE} & MDE\\
\cline{3-11}
& & center & range & center & range & center & range & center & range & interval\\
% data set C1
\midrule
\multirow{5}{*}{\tabincell{c}{C1 \\ (ResNet-18)} } & SVR &0.0209 & 0.0867 & 0.1149 & 0.2544 & 0.2241 & 1.5644 & 0.2074 & 0.5813 & 0.5472\\
& RF &0.0199 & 0.0855 & 0.1111 & 0.2546 & 0.2192 & 2.0910 & 0.2005 & 0.5824 & 0.5456\\
& LSTM &0.0208 & 0.0846 & 0.1149 & 0.2536 & 0.2182 & 2.3309 & 0.2076 & 0.5797 & 0.5463\\
& XGBoost &0.0248 & 0.1033 & 0.1248 & 0.2666 & 0.2453 & 1.8905 & 0.2261 & 0.6072 & 0.5648\\
& WNN &0.0199 & 0.0857 & 0.1109 & 0.2548 & 0.2180 & 1.9981 & 0.1994 & 0.5812 & 10.3939\\
\midrule
% data set C2
\multirow{5}{*}{\tabincell{c}{C2\\ (ResNet-34)} } & SVR &0.0295 & 0.092 & 0.1398 & 0.2643 & 0.4020 & 3.2739 & 0.2872 & 0.6075 & 0.5634\\
& RF &0.0287 & 0.0875 & 0.1361 & 0.2592 & 0.4069 & 3.0657 & 0.2794 & 0.6010 & 0.5585\\
& LSTM &0.0292 & 0.0865 & 0.1379 & 0.2582 & 0.4088 & 2.9194 & 0.2828 & 0.6035 & 0.5582\\
& XGBoost &0.0320 & 0.0917	 & 0.1414 & 0.2617 & 0.4217 & 3.1843 & 0.2887 & 0.6030 & 0.5657\\
& WNN &0.0289 & 0.0869 & 0.1369 & 0.2584 & 0.4079 & 3.0666 & 0.2811 & 0.5995 & 10.651\\
\midrule
% data set C3
\multirow{5}{*}{\tabincell{c}{C3\\ (ResNet-50)}} & SVR &0.0264 & 0.0898 & 0.1303 & 0.2611 & 0.8878 & 4.6571 & 0.2733 & 0.5974 & 0.5596\\
& RF &0.0260 & 0.0869 & 0.1301 & 0.2589 & 0.9981 & 4.5655 & 0.273 & 0.5944 & 0.5574\\
& LSTM &0.0268 & 0.0862 & 0.1325 & 0.2576 & 1.0910 &6.981 & 0.277 & 0.5912 & 0.5577\\
& XGBoost &0.0282 & 0.0861 & 0.1345 & 0.2576 & 0.7077 & 4.5309 & 0.2823 & 0.5924 & 0.5600\\
& WNN &0.0258 & 0.0866 & 0.1300 & 0.2585 & 0.4431 & 2.8601 & 0.2723 & 0.5926 & 10.6573\\
\midrule
% data set SP500
\multirow{5}{*}{\tabincell{c}{S\&P500 \\ (ResNet-50)}} & SVR &0.1787 & 0.0193 & 0.3923 & 0.1072 & 0.5826 & 1.2559 & 0.8362 & 0.6304 & 0.6470\\
& RF &0.1932 & 0.0275 & 0.4110 & 0.1056 & 0.6132 & 0.5431 & 0.8959 & 0.7078 & 0.6660\\
& LSTM &0.187 & 0.0286 & 0.4035 & 0.1092 & 0.6012 & 0.5506 & 0.8714 & 0.7488 & 0.6615\\
& XGBoost &0.1968 & 0.0274 & 0.4097 & 0.1058 & 0.6108 & 0.5628 & 0.9087 & 0.7223 & 0.6657\\
& WNN &0.1923 & 0.0276 & 0.4100 & 0.1060 & 0.6116 & 0.5483 & 0.8925 & 0.7153 & 17.9001\\
\bottomrule
\end{tabular}
\end{table}

\begin{table}[H]
\footnotesize
\centering
\setlength{\abovecaptionskip}{0pt}%
\setlength{\belowcaptionskip}{3pt}%
\caption{\normalsize The prediction performance of the models on features extracted from the GADF representation}\label{fenresulsD}
\begin{tabular}{clccccccccc}
\toprule
\multirow{2}{*}{\tabincell{c}{Data sets\\ (FEN)}}&\multirow{2}{*}{Models}&\multicolumn{2}{c}{MSE}  & \multicolumn{2}{c}{MAE} & \multicolumn{2}{c}{MAPE} 
& \multicolumn{2}{c}{SMAPE} & MDE\\
\cline{3-11}
& & center & range & center & range & center & range & center & range & interval\\
% data set C1
\midrule
\multirow{5}{*}{\tabincell{c}{C1 \\ (ResNet-18)} } & SVR &0.0211 & 0.0865 & 0.1158 & 0.2532 & 0.2249 &0.6679 & 0.2092 & 0.5807 & 0.5466\\
& RF &0.0199 & 0.0843 & 0.1111 & 0.2525 & 0.2184 & 0.5691 & 0.2005 & 0.5778 & 0.5433\\
& LSTM &0.0203 & 0.0846 & 0.1127 & 0.2536 & 0.2186 & 0.4479 & 0.2036 & 0.5796 & 0.5455	\\
& XGBoost &0.0249 & 0.0891 & 0.1243 & 0.2543 & 0.2432 & 0.7879 & 0.2238 & 0.5787 & 0.5523\\
& WNN &0.0199 & 0.0850 & 0.1108 & 0.2543 & 0.2177 & 0.3346 & 0.1996 & 0.5802 & 10.3892\\
\midrule
% data set C2
\multirow{5}{*}{\tabincell{c}{C2\\ (ResNet-34)} } & SVR &0.0305 & 0.0903 & 0.1413 & 0.2618 & 0.4087 & 3.0854 & 0.2897 & 0.6039 & 0.5637\\
& RF &0.0289 & 0.0869 & 0.1375 & 0.2580 & 0.4092 & 3.0452 & 0.2821 & 0.5988 & 0.5580\\
& LSTM &0.0291 & 0.0883 & 0.1375 & 0.2592 & 0.4098 & 3.2188 & 0.2820 & 0.5959 & 0.5595\\
& XGBoost &0.0345 & 0.0991 & 0.1509 & 0.2734 & 0.4297 & 3.0327 & 0.3075 & 0.6277 & 0.5786\\
& WNN &0.0294 & 0.0872 & 0.1382 & 0.2582 & 0.4077 & 3.0717 & 0.2812 & 0.5997 & 10.6550\\
\midrule
% data set C3
\multirow{5}{*}{\tabincell{c}{C3\\ (ResNet-50)}} & SVR &0.0266 & 0.0883 & 0.1313 & 0.2591 & 0.4030 & 0.6673 & 0.2754 & 0.5963 & 0.5580\\
& RF &0.0258 & 0.0866 & 0.1296 & 0.2577 & 0.3901 & 0.6766 & 0.2721 & 0.5924 & 0.5563\\
& LSTM &0.0262 & 0.0865 & 0.1307 & 0.2584 & 0.3419 & 0.7801 & 0.2738 & 0.5963 & 0.5576\\
& XGBoost &0.0309 & 0.1052 & 0.1414 & 0.2711 & 0.6681 & 0.7079 & 0.2953 & 0.6173 & 0.5746\\
& WNN &0.0260 & 0.0860 & 0.1303 & 0.2575 & 0.7971 & 0.9012 & 0.2726 & 0.5926 & 10.6563\\
\midrule
% data set SP500
\multirow{5}{*}{\tabincell{c}{S\&P500 \\ (ResNet-50)}} & SVR &0.1795 & 0.0196 & 0.3936 & 0.1093 & 0.5849 & 1.2912 & 0.8403 & 0.6384 & 0.6482\\
& RF &0.1929 & 0.0275 & 0.4106 & 0.1058 & 0.6126 & 0.5439 & 0.8945 & 0.71 & 0.6656\\
& LSTM &0.2060 & 0.0283 & 0.4264 & 0.1082 & 0.6384 & 0.5476 & 0.9480 & 0.7373 & 0.6773\\
& XGBoost &0.1963 & 0.0271 & 0.4092 & 0.1042 & 0.6103 & 0.5407 & 0.9072 & 0.6948 & 0.6648\\
& WNN &0.1936 & 0.0275 & 0.4112 & 0.1057 & 0.6137 & 0.5439 & 0.8971 & 0.7079 & 17.9221\\
\bottomrule
\end{tabular}
\end{table}

\begin{table}[H]
\footnotesize
\centering
\setlength{\abovecaptionskip}{0pt}%
\setlength{\belowcaptionskip}{3pt}%
\caption{\normalsize The prediction performance of the models on features extracted from the MTF representation}\label{fenresulsM}
\begin{tabular}{clccccccccc}
\toprule
\multirow{2}{*}{\tabincell{c}{Data sets\\ (FEN)}}&\multirow{2}{*}{Models}&\multicolumn{2}{c}{MSE}  & \multicolumn{2}{c}{MAE} & \multicolumn{2}{c}{MAPE} 
& \multicolumn{2}{c}{SMAPE} & MDE\\
\cline{3-11}
& & center & range & center & range & center & range & center & range & interval\\
% data set C1
\midrule
\multirow{5}{*}{\tabincell{c}{C1 \\ (ResNet-18)} } & SVR &0.0218 & 0.0895 & 0.1170 & 0.2577 & 0.2284 & 0.3433 & 0.2113 & 0.5864 & 0.5518\\
& RF &0.0198 & 0.0854 & 0.1109 & 0.2546 & 0.2185 & 0.3019 & 0.2001 & 0.5820 & 0.5451\\
& LSTM &0.0199 & 0.0851 & 0.1108 & 0.2545 & 0.2170 & 0.3435 & 0.2001 & 0.5827 & 0.5451\\
& XGBoost &0.0252 & 0.1036 & 0.1251 & 0.2704 & 0.2451 & 0.3601 & 0.2251 & 0.6153 & 0.5669\\
& WNN &0.0197 & 0.0854 & 0.1106 & 0.2550 & 0.2176 & 0.3301 & 0.1995 & 0.5809 & 10.3911\\
\midrule
% data set C2
\multirow{5}{*}{\tabincell{c}{C2\\ (ResNet-34)} } & SVR &0.0294 & 0.092 & 0.1387 & 0.2631 & 0.3992 & 3.1076 & 0.2853 & 0.6064 & 0.5635\\
& RF &0.0292 & 0.0878 & 0.1376 & 0.2593 & 0.4092 & 3.0432 & 0.2823 & 0.6014 & 0.5594\\
& LSTM &0.0290 & 0.0885 & 0.1371 & 0.2601 & 0.4084 & 2.6919 & 0.2814 & 0.6138 & 0.5596\\
& XGBoost &0.0326 & 0.1047 & 0.1436 & 0.2743 & 0.4242 & 3.0476 & 0.2929 & 0.6262 & 0.5759\\
& WNN &0.0289 & 0.0875 & 0.1370 & 0.2589 & 0.4069 & 3.0609 & 0.2808 & 0.5997 & 10.6477\\
\midrule
% data set C3
\multirow{5}{*}{\tabincell{c}{C3\\ (ResNet-50)}} & SVR &0.0266 & 0.0882 & 0.1337 & 0.2605 & 0.6601 & 2.4581 & 0.2804 & 0.5968 & 0.5593\\
& RF &0.0257 & 0.0865 & 0.1294 & 0.2580 & 0.6969 & 4.3201 & 0.2720 & 0.5929 & 0.5562\\
& LSTM &0.0260 & 0.0868 & 0.1309 & 0.2586 & 0.4749 & 2.9044 & 0.2744 & 0.5936 & 0.5575\\
& XGBoost &0.0301 & 0.1086 & 0.1407 & 0.2808 & 0.5601 & 3.0901 & 0.2932 & 0.6313 & 0.5796\\
& WNN &0.0258 & 0.0866 & 0.1300 & 0.2588 & 0.4601 & 2.9001 & 0.2724 & 0.5928 & 10.6573\\
\midrule
% data set SP500
\multirow{5}{*}{\tabincell{c}{S\&P500 \\ (ResNet-50)}} & SVR &0.1765 & 0.0191 & 0.3899 & 0.1062 & 0.5792 & 1.2381 & 0.8290 & 0.6260 & 0.6451\\
& RF &0.1933 & 0.0275 & 0.4111 & 0.1059 & 0.6134 & 0.5430 & 0.8961 & 0.7105 & 0.6660\\
& LSTM &0.1919 & 0.0269 & 0.4096 & 0.1041 & 0.6110 & 0.5417 & 0.8912 & 0.6907 & 0.6645\\
& XGBoost &0.1987 & 0.0279 & 0.4124 & 0.1076 & 0.6141 & 0.5747 & 0.9123 & 0.7459 & 0.6679\\
& WNN &0.1934 & 0.0276 & 0.4113 & 0.1059 & 0.6138 & 0.5442 & 0.8969 & 0.7103 & 17.9248\\
\bottomrule
\end{tabular}
\end{table}

Compared to the prediction results of the raw data presented in Table \ref{originresults}, the results obtained through imaging and feature extraction (Tables \ref{fenresulsR}, \ref{fenresulsS}, \ref{fenresulsD}, and \ref{fenresulsM}) demonstrate significant improvements across all performance metrics. This improvement is evident for various models, imaging approaches, and DGPs. For instance, let's consider the LSTM model's performance metric on DGP1(C1) before and after feature extraction from the RP representation. Prior to feature extraction, the metric values are 4.2308, 33.5927, 1.6183, 5.0193, 1.8027, 0.1285, 1.0579, 0.1263, and 2.3575. However, after feature extraction, these values are reduced to 0.0199, 0.0853, 0.1114, 0.2541, 0.2200, 2.4456, 0.2010, 0.5803, and 0.5451. This indicates a substantial improvement in the performance metrics, and similar improvements can be observed for other imaging approaches, models, and DGPs. The similarity between our prediction results based on raw data sets and the experimental results of \textcite{xu2021novel} validates the necessity of proper feature extraction before conducting time series prediction. The balanced prediction performance observed after feature extraction signifies the effectiveness of the feature extraction procedure in improving the predictive capabilities of the models. These results highlight the importance of leveraging informative features extracted from the time series data to achieve more desirable and accurate predictions.

The results obtained in our study validate the effectiveness of the proposed feature extraction procedure. They demonstrate that by extracting informative features from the raw data, significant improvements in prediction performance can be achieved. This underscores the importance of feature extraction in capturing the underlying patterns and dynamics of the time series data. Furthermore, the results suggest that once a favorable representation of the raw data is obtained through proper feature extraction, the selection of the prediction model has minimal impact on the prediction performance. In other words, the choice of model becomes less influential in determining the accuracy of predictions when informative features are extracted from the data.

\subsection{State-of-the-art time series prediction results based on neural networks}\label{otherdeep}

In the previous subsections, the effectiveness of the feature extraction procedure was validated in terms of improving the predictive performance of statistical learning models after feature extraction. However, it did not compare the performance of end-to-end neural network-based time series prediction models. In this subsection, we address this gap by selecting state-of-the-art (SOTA) time series forecasting models based on neural networks and comparing their performance with the results presented in Subsection \ref{fenp}. The selected SOTA models for comparison include: N-BEATS \cite{DBLP:conf/iclr/OreshkinCCB20}, N-HiTS \cite{DBLP:journals/corr/abs-2201-12886}
TCN \cite{DBLP:journals/corr/abs-1803-01271}, Transformer \cite{DBLP:conf/nips/VaswaniSPUJGKP17}. The subsequent table presents the prediction performance of these SOTA models on the four DGPs. By comparing their performance with the results obtained from the feature extraction-based models, we can assess the relative performance and advantages of the SOTA models in time series prediction. 

\begin{table}[H]
\footnotesize
\centering 
\caption{\normalsize Raw data sets prediction results based on some current SOTA structure} \label{deepneural}
\setlength{\tabcolsep}{1.2mm}{  
\begin{tabular}{clccccccccc}
\toprule
\multirow{2}{*}{\tabincell{c}{Data sets}}&\multirow{2}{*}{Models}&\multicolumn{2}{c}{MSE}  & \multicolumn{2}{c}{MAE} & \multicolumn{2}{c}{MAPE} 
& \multicolumn{2}{c}{SMAPE} & MDE\\
\cline{3-11}
& & center & range & center & range & center & range & center & range & interval\\ 
% data set C1
\midrule
\multirow{4}{*}{C1} 
& N-BEATS& 9.9190 & 47.112 & 2.4949 & 5.9344 & 151.61 & 14.851 & 152.17 & 14.919 & 2.6192 \\
& N-HiTS  &10.723 & 50.792 & 2.5168 & 5.9653 & 149.84 & 14.510 & 147.65 & 15.116 & 2.6376\\
& TCN  & 9.4391 & 53.541 & 2.4252 & 6.1267 & 105.02 & 15.419 & 172.77 & 15.446  & 2.6487 \\
& Transformer & 9.5480 & 34.886 & 2.4183 & 5.1827 & 134.80 & 13.248 & 143.28 & 13.018 & 2.4777\\
\midrule	

\multirow{4}{*}{C2} 
& N-BEATS & 19.415 & 1.8316 & 3.4549 & 1.1626 & 187.35 & 51.571 & 126.29  & 40.900 & 1.9560	\\
& N-HiTS  &  22.541 & 1.8456 & 3.8845 & 1.1485 & 337.77 & 46.387 & 138.38 & 40.501& 2.0486\\
& TCN  & 14.828 & 1.4676 & 3.1268 & 1.0880 & 237.84& 48.709 & 120.03 & 38.702 & 1.8530\\
& Transformer  &  15.217 & 1.5012 & 3.1774 & 1.0879& 316.35 & 48.776& 117.77 & 38.662 & 1.8679 \\
\midrule

\multirow{4}{*}{C3}
& N-BEATS &1.0815 & 281.85  & 0.8244 & 14.156  & 1.7569& 32.054 & 1.7626 & 33.231 & 3.7717\\
& N-HiTS  & 0.9847 & 331.93  & 0.8116  & 15.2450 & 1.7234 & 37.086 & 1.7353 & 35.915 & 3.9136 \\
& TCN  & 1.8454  & 231.39& 1.1044 & 13.0262 & 2.3645 & 33.991 & 2.3499 & 30.066 & 3.6314\\
& Transformer & 0.86176 & 220.50 & 0.7782 & 12.882 & 1.6679 & 34.750 & 1.6629 & 29.713 & 3.5995 \\

%\midrule
%\multirow{4}{*}{S\&P500} 
%& N-BEATS &&&&&&&&&\\
%& N-HiTS  &&&&&&&&&\\
%& TCN  &&&&&&&&&\\
%& Transformer  &&&&&&&&& \\
\bottomrule
\end{tabular}
}
\end{table}

It is important to note that the prediction results for the S\&P500 index are not reported in the previous tables because the selected SOTA structures are not suitable for this particular dataset. As discussed in Subsection \ref{fenp}, the SOTA models may not be effective for the S\&P500 index data. Comparing Table \ref{deepneural} with either Table \ref{fenresulsR}, \ref{fenresulsS}, \ref{fenresulsD}, or \ref{fenresulsM}, it becomes apparent that the results across these tables are remarkably similar. Therefore, it is sufficient to focus on the comparison between Table \ref{fenresulsR} and Table \ref{deepneural}. Upon comparing these tables, it is evident that the feature extraction procedure outperforms the selected SOTA structures on the three DGPs. This finding demonstrates that the feature extraction procedure is able to yield a better representation of the raw dataset than the sophisticated network structures employed in the SOTA models.

The superior performance of the feature extraction procedure suggests that it captures and extracts relevant information and features from the time series data more effectively than the SOTA structures. This emphasizes the importance of proper feature extraction in time series analysis and prediction tasks, highlighting its ability to enhance the representation of the data and improve prediction accuracy.

\section{Conclusion}\label{sec5}

In this paper, we introduced a novel approach for time series prediction by combining time series imaging and feature-based transfer learning. The proposed approach serves as a self-contained time series feature extraction procedure. Through simulation experiments and real data studies, we demonstrate that the effectiveness of the proposed feature extraction procedure is independent of the specific prediction model used. By applying the feature extraction procedure to various prediction models, we observe a significant improvement in their performance. Importantly, we find that the performance of different models becomes highly similar after feature extraction. This suggests that the feature extraction procedure itself plays a crucial role in enhancing the predictive capabilities of the models, making the specific choice of prediction model less influential.

Additionally, our experiments revealed that the choice of imaging approach had a minimal impact on the prediction performance. This finding suggests that a single imaging approach can be utilized for real data studies, thereby reducing the computational effort required for feature extraction.  Furthermore, despite the use of a short segmentation length that results in moderately large images, we were able to effectively extract features from these images. This success can be attributed to the rapid advancements in computer vision technology, which have enabled efficient and accurate feature extraction even from relatively large images. Consequently, the proposed feature extraction procedure remains applicable to interval-valued time series of moderate size.

In conclusion, there are two potential future research directions for this study. Firstly, while we have utilized ResNet models of different depths as candidates for the FENs, it is possible that other network structures may be more suitable when combined with the feature extraction procedure. Thus, future exploration could involve investigating different network architectures that can effectively extract features from time series data. Secondly, the feature extraction procedure proposed in this study focuses on univariate time series imaging approaches. However, in the future, it could be valuable to develop a feature extraction procedure based on a multivariate time series imaging method, such as joint recurrence plots \cite{ROMANO2004214}. This approach could be applied to interval vector-valued time series prediction, where multiple variables are considered together.

By pursuing these future research directions, we can further enhance the feature extraction process and explore its applicability to a wider range of time series data, including multivariate and interval vector-valued time series. These advancements will contribute to the development of more effective and comprehensive approaches for time series prediction and analysis.

\section*{Acknowledgments}
\addcontentsline{toc}{section}{Acknowledgments}
The research work described in this paper was supported by the National Natural Science Foundation of China (Nos. 72071008) and and Beijing Natural Science Foundation (Nos. Z210003).

%\bibliography{./bib/reference.bib}
%\bibliographystyle{plain}

\printbibliography

@article{xu2021novel,
	title={A novel hybrid {ARIMA} and regression tree model for the interval-valued time series},
	author={Xu, Min and Qin, Zhongfeng},
	journal={Journal of Statistical Computation and Simulation},
	volume={91},
	number={5},
	pages={1000--1015},
	year={2021},
	publisher={Taylor \& Francis}
}

@conference{arroyo2006introducing,
	title={Introducing interval time series: Accuracy measures},
	author={Arroyo, Javier and Mat{\'e}, Carlos},
	booktitle={COMPSTAT 2006, proceedings in computational statistics},
	pages={1139--1146},
	year={2006},
	publisher={Physica-Verlag Heidelberg}
}

@article{san2007imlp,
	title={{iMLP: Applying multi-layer perceptrons to interval-valued data}},
	author={San Roque, Antonio Mu{\~n}oz and Mat{\'e}, Carlos and Arroyo, Javier and Sarabia, {\'A}ngel},
	journal={Neural Processing Letters},
	volume={25},
	number={2},
	pages={157--169},
	year={2007},
	publisher={Springer}
}

@inproceedings{arroyo2007exponential,
	title={Exponential smoothing methods for interval time series},
	author={Arroyo, Javier and San Roque, A Mu{\~n}oz and Mat{\'e}, Carlos and Sarabia, A},
	booktitle={Proceedings of the 1st European Symposium on Time Series Prediction},
	pages={231--240},
	year={2007}
}

@book{fan2008nonlinear,
	title={Nonlinear time series: nonparametric and parametric methods},
	author={Fan, Jianqing and Yao, Qiwei},
	year={2008},
	publisher={Springer Science \& Business Media}
}

@article{neto2008centre,
	title={Centre and range method for fitting a linear regression model to symbolic interval data},
	author={Neto, Eufr{\'a}sio de A Lima and de Carvalho, Francisco de AT},
	journal={Computational Statistics \& Data Analysis},
	volume={52},
	number={3},
	pages={1500--1515},
	year={2008},
	publisher={Elsevier}
}

@article{maia2008forecasting,
	title={Forecasting models for interval-valued time series},
	author={Maia, Andr{\'e} Luis S and de Carvalho, Francisco de AT and Ludermir, Teresa B},
	journal={Neurocomputing},
	volume={71},
	number={16-18},
	pages={3344--3352},
	year={2008},
	publisher={Elsevier}
}

@inproceedings{he2016deep,
	title={Deep residual learning for image recognition},
	author={He, Kaiming and Zhang, Xiangyu and Ren, Shaoqing and Sun, Jian},
	booktitle={Proceedings of the IEEE conference on computer vision and pattern recognition},
	pages={770--778},
	year={2016}
}

@book{Goodfellow-et-al-2016,
	title={Deep Learning},
	author={Ian Goodfellow and Yoshua Bengio and Aaron Courville},
	publisher={MIT Press},
	year={2016}
}

@article{HochreiterSepp1997,
	author = {Hochreiter, Sepp and Schmidhuber, Jürgen},
	year = {1997},
	month = {12},
	pages = {1735-1780},
	title = {{Long Short-term Memory}},
	volume = {9},
	journal = {Neural Computation}
}

@article{CutlerAdele2011,
	author = {Breiman, Leo},
	title = {{Random Forests}},
	year = {2001},
	issue_date = {October 1 2001},
	publisher = {Kluwer Academic Publishers},
	address = {USA},
	volume = {45},
	number = {1},
	issn = {0885-6125},
	journal = {Machine Learning},
	pages = {5–32},
	numpages = {28}
}

@article{AlexandridisAntonios2011,
	title = {Wavelet neural networks: A practical guide},
	journal = {Neural Networks},
	volume = {42},
	pages = {1-27},
	year = {2013},
	issn = {0893-6080},
	author = {Antonios K. Alexandridis and Achilleas D. Zapranis}
}

@article{SmolaAlextutorialsvr2004,
	author = {Smola, Alex and Schölkop, Bernhard},
	year = {2004},
	month = {08},
	pages = {199-222},
	title = {A tutorial on support vector regression},
	volume = {14},
	journal = {Statistics and Computing},
}

@inproceedings{DBLP:conf/kdd/ChenG16,
	author    = {Tianqi Chen and
	Carlos Guestrin},
	editor    = {Balaji Krishnapuram and
	Mohak Shah and
	Alexander J. Smola and
	Charu C. Aggarwal and
	Dou Shen and
	Rajeev Rastogi},
	title     = {{XGBoost: {A} Scalable Tree Boosting System}},
	booktitle = {Proceedings of the 22nd {ACM} {SIGKDD} International Conference on
	Knowledge Discovery and Data Mining},
	pages     = {785--794},
	publisher = {{ACM}},
	year      = {2016},
	bibsource = {dblp computer science bibliography, https://dblp.org}
}

@article{eckmann1995recurrence,
	title={Recurrence plots of dynamical systems},
	author={Eckmann, Jean-Pierre and Kamphorst, S Oliffson and Ruelle, David and others},
	journal={World Scientific Series on Nonlinear Science Series A},
	volume={16},
	pages={441--446},
	year={1995},
	publisher={WORLD SCIENTIFIC PUBLISHING}
}

@inproceedings{wang2015encoding,
	title={Encoding time series as images for visual inspection and classification using tiled convolutional neural networks},
	author={Wang, Zhiguang and Oates, Tim},
	booktitle={Workshops at the twenty-ninth AAAI conference on artificial intelligence},
	year={2015}
}

@article{tian2021drought,
	title={{Drought prediction based on feature-based transfer learning and time series imaging}},
	author={Tian, Wan and Wu, Jiujing and Cui, Hengjian and Hu, Tao},
	journal={IEEE Access},
	volume={9},
	pages={101454--101468},
	year={2021},
	publisher={IEEE}
}

@article{ROMANO2004214,
	author = {M. Carmen Romano and Marco Thiel and Jürgen Kurths and Werner {von Bloh}},
	title = {Multivariate recurrence plots},
	journal = {Physics Letters A},
	volume = {330},
	number = {3},
	pages = {214-223},
	year = {2004},
	issn = {0375-9601}
}

@incollection{han2016vector,
	title={A vector autoregressive moving average model for interval-valued time series data},
	author={Han, Ai and Hong, Yongmiao and Wang, Shouyang and Yun, Xin},
	booktitle={Essays in Honor of Aman Ullah},
	year={2016},
	publisher={Emerald Group Publishing Limited}
}

@article{he2009impacts,
	title={Impacts of interval computing on stock market variability forecasting},
	author={He, Ling T and Hu, Chenyi},
	journal={Computational Economics},
	volume={33},
	number={3},
	pages={263--276},
	year={2009},
	publisher={Springer}
}

@article{gonzalez2013constrained,
	title={Constrained regression for interval-valued data},
	author={Gonz{\'a}lez-Rivera, Gloria and Lin, Wei},
	journal={Journal of Business \& Economic Statistics},
	volume={31},
	number={4},
	pages={473--490},
	year={2013},
	publisher={Taylor \& Francis}
}

@inproceedings{han2012autoregressive,
	title={Autoregressive conditional models for interval-valued time series data},
	author={Han, Ai and Hong, Yongmiao and Wang, Shouyang},
	booktitle={The 3rd International Conference on Singular Spectrum Analysis and Its Applications},
	pages={27},
	year={2012}
}

@article{Sun2022,
	doi = {10.1016/j.ejor.2021.11.015},
	url = {https://doi.org/10.1016/j.ejor.2021.11.015},
	year = {2022},
	month = sep,
	publisher = {Elsevier {BV}},
	volume = {301},
	number = {2},
	pages = {772--784},
	author = {Yuying Sun and Xinyu Zhang and Alan T.K. Wan and Shouyang Wang},
	title = {Model averaging for interval-valued data},
	journal = {European Journal of Operational Research}
}

@book{moore1966interval,
	title={Interval analysis},
	author={Moore, Ramon E},
	volume={4},
	year={1966},
	publisher={Prentice-Hall Englewood Cliffs}
}

@inproceedings{fawaz2018transfer,
	title={Transfer learning for time series classification},
	author={Fawaz, Hassan Ismail and Forestier, Germain and Weber, Jonathan and Idoumghar, Lhassane and Muller, Pierre-Alain},
	booktitle={2018 IEEE international conference on big data (Big Data)},
	pages={1367--1376},
	year={2018},
	organization={IEEE}
}

@article{ye2018novel,
	title={A novel transfer learning framework for time series forecasting},
	author={Ye, Rui and Dai, Qun},
	journal={Knowledge-Based Systems},
	volume={156},
	pages={74--99},
	year={2018},
	publisher={Elsevier}
}

@inproceedings{vercruyssen2017transfer,
	title={Transfer learning for time series anomaly detection},
	author={Vercruyssen, Vincent and Meert, Wannes and Davis, Jesse},
	booktitle={Proceedings of the Workshop and Tutorial on Interactive Adaptive Learning@ ECMLPKDD 2017},
	volume={1924},
	pages={27--37},
	year={2017},
	organization={CEUR Workshop Proceedings}
}

@inproceedings{DBLP:conf/iclr/OreshkinCCB20,
	author    = {Boris N. Oreshkin and
	Dmitri Carpov and
	Nicolas Chapados and
	Yoshua Bengio},
	title     = {{N-BEATS:} Neural basis expansion analysis for interpretable time
	series forecasting},
	booktitle = {8th International Conference on Learning Representations, {ICLR} 2020,
	Addis Ababa, Ethiopia, April 26-30, 2020},
	publisher = {OpenReview.net},
	year      = {2020},
	url       = {https://openreview.net/forum?id=r1ecqn4YwB},
	bibsource = {dblp computer science bibliography, https://dblp.org}
}

@article{DBLP:journals/corr/abs-2201-12886,
	author    = {Cristian Challu and
	Kin G. Olivares and
	Boris N. Oreshkin and
	Federico Garza and
	Max Mergenthaler and
	Artur Dubrawski},
	title     = {{N-HiTS: Neural Hierarchical Interpolation for Time Series Forecasting}},
	journal   = {CoRR},
	volume    = {abs/2201.12886},
	year      = {2022},
	url       = {https://arxiv.org/abs/2201.12886},
	eprinttype = {arXiv},
	eprint    = {2201.12886},
	bibsource = {dblp computer science bibliography, https://dblp.org}
}

@article{DBLP:journals/corr/abs-1803-01271,
	author    = {Shaojie Bai and
	J. Zico Kolter and
	Vladlen Koltun},
	title     = {{An Empirical Evaluation of Generic Convolutional and Recurrent Networks
	for Sequence Modeling}},
	journal   = {CoRR},
	volume    = {abs/1803.01271},
	year      = {2018},
	url       = {http://arxiv.org/abs/1803.01271},
	eprinttype = {arXiv},
	eprint    = {1803.01271},
	bibsource = {dblp computer science bibliography, https://dblp.org}
}

@inproceedings{DBLP:conf/nips/VaswaniSPUJGKP17,
	author    = {Ashish Vaswani and
	Noam Shazeer and
	Niki Parmar and
	Jakob Uszkoreit and
	Llion Jones and
	Aidan N. Gomez and
	Lukasz Kaiser and
	Illia Polosukhin},
	editor    = {Isabelle Guyon and
	Ulrike von Luxburg and
	Samy Bengio and
	Hanna M. Wallach and
	Rob Fergus and
	S. V. N. Vishwanathan and
	Roman Garnett},
	title     = {{Attention is All you Need}},
	booktitle = {Advances in Neural Information Processing Systems 30: Annual Conference
	on Neural Information Processing Systems 2017, December 4-9, 2017,
	Long Beach, CA, {USA}},
	pages     = {5998--6008},
	year      = {2017},
	url       = {https://proceedings.neurips.cc/paper/2017/hash/3f5ee243547dee91fbd053c1c4a845aa-Abstract.html},
	bibsource = {dblp computer science bibliography, https://dblp.org}
}

@article{arroyo2011different,
	title={Different approaches to forecast interval time series: a comparison in finance},
	author={Arroyo, Javier and Esp{\'\i}nola, Rosa and Mat{\'e}, Carlos},
	journal={Computational Economics},
	volume={37},
	pages={169--191},
	year={2011},
	publisher={Springer}
}
\end{document}